# High-pressure characterization of multifunctional $CrVO_4$

P. Botella[1,*], S. López-Moreno[2], D. Errandonea[3], F. J. Manjón[4], J. A. Sans[4], D. Vie[5], and A. Vomiero[1,6]

*1. Division of Materials Science, Department of Engineering Sciences and Mathematics, Luleå University of Technology, SE-97187 Luleå, Sweden*

*2. CONACYT – División de Materiales Avanzados, IPICYT, Camino a la Presa San José 2055, San Luis Potosí, S.L.P. 78216, México*

*3. Departament de Física Aplicada-ICMUV, MALTA Consolider Team, Universitat de València, Dr. Moliner 50, 46100 Burjassot, Spain*

*4. Instituto de Diseño para la Fabricación y Producción Automatizada, MALTA Consolider Team, Universitat Politècnica de València, Camí de Vera s/n, 46022 València, Spain*

*5. Institut de Ciència dels Materials de la Universitat de València, Apartado de Correos 2085, E-46071 València, Spain.*

*6. Department of Molecular Sciences and Nanosystems, Ca' Foscari University of Venice, via Torino 155, 30172 Venezia, Italy*

**E-mail: pablo.botella.vives@ltu.se*

## Abstract

The structural stability and physical properties of $CrVO_4$ under compression were studied by X-ray diffraction, Raman spectroscopy, optical absorption, resistivity measurements, and *ab initio* calculations up to 10 GPa. High-pressure X-ray diffraction and Raman measurements show that $CrVO_4$ undergoes a phase transition from the ambient pressure orthorhombic $CrVO_4$-type structure (*Cmcm* space group, phase III) to the high-pressure monoclinic $CrVO_4$-V phase, which is isomorphic to the wolframite structure. Such a phase transition ($CrVO_4$-type $\rightarrow$ wolframite), driven by pressure, also was previously observed in indium vanadate. The crystal structure of both phases and the pressure dependence in unit-cell parameters, Raman-active modes, resistivity, and electronic band gap, is reported. Vanadium atoms are sixth-fold coordinated in the wolframite phase, which is related to the collapse in the volume at the phase transition. Besides, we also observed drastic changes in the phonon spectrum, a drop of the band-gap, and a sharp decrease of resistivity. All the observed phenomena are explained with the help of first-principles calculations.

## INTRODUCTION

Among transition metal orthovanadates of the type $A^{3+}VO_4$ ($A$ = Cr, In, Fe, Tl, Al, Bi, etc.), chromium vanadate ($CrVO_4$) is one of the prototype compounds. In addition, its orthorhombic crystal structure is the model used to describe a mineral family with an isomorphic structure, which includes orthophosphates and orthosilicates among others, like some recently proposed orthoborates [1][2]. $CrVO_4$-type compounds are of special interest due to their multifunctionality. Among its applications, the possibility to act as superprotonic conductors [3], catalytic materials [4], or as cathode for lithium-ion batteries [5] are the most remarkable. Specifically, $CrVO_4$, apart from its interesting structural and magnetic properties [6][7], possesses a band-gap in the visible region, which is ideal for photocatalytic and electrochemical applications [8][9][10].

The properties of $CrVO_4$ (and therefore its applications) are intimately related to its crystal structure (arrangement of the atoms in the solid), which define the electronic band structure as well as the magnetic and vibrational properties of the material. Specifically, in $A^{3+}VO_4$ vanadate-based ternary metal oxides, in which V is in tetrahedral coordination, the valence band maximum (VBM) and conduction band minimum (CBM), are mainly dominated by O 2p and V 3d orbitals, with a small contribution of Cr 3d electrons. The partial substitution of Cr by a different $A^{3+}$ atom, or the application of pressure are two methods of modifying the energy gap and the magnetic properties of $CrVO_4$ [7][10], tuning this way its properties for specific applications.

Chromium vanadate crystallizes in three different polymorphs viz. orthorhombic, monoclinic, and tetragonal. The metastable monoclinic $CrVO_4$-I [$\alpha$-$MnMoO_4$-type structure, space group (SG): *C*2/*m*, No. 12, $Z$ = 8] has been synthesized by different methods [11]. Another monoclinic system ($CrVO_4$-IV, SG: *Cmm2*, No. 35) was also synthetized by soft chemistry at 450 °C [12]. The orthorhombic $CrVO_4$-III phase ($CrVO_4$-type structure, SG: *Cmcm*, No. 63, $Z$ = 4) is the most stable and the prototype structure [9][11][13]. Finally, the metastable tetragonal phase (rutile-type structure, SG: $P4_2/mnm$, No. 136, $Z$ = 1) can be only synthetized under high-temperature and high-pressure conditions [14][15][16]. Traditionally, the polymorphs of $CrVO_4$ have been labeled following the notation used for indium vanadate ($InVO_4$) due to their structural similarities. The monoclinic form, $CrVO_4$-I, is isomorphic to $InVO_4$-I and the orthorhombic form, $CrVO_4$-III, is isostructural to $InVO_4$-III, as well as to $TlVO_4$ and $FeVO_4$ [14][17][18][19][20]. The $CrVO_4$-II and $CrVO_4$-IV phases; however, bear no resemblance with $InVO_4$ polymorphs.

In the last decade, special effort has been put in the study of ternary oxides under pressure and, specifically, to the behavior of orthovanadates [21][22]. $CrVO_4$-type compounds are interesting candidates to be studied under pressure since the field map diagram [23] locate them as intermediate structure between quartz-like structures with four-fold coordinated cations and structures with six-fold coordinated cations [24]; for instance, wolframites [25]. $InVO_4$-III has been studied under pressure at room temperature only recently [26][27][28]. Interesting phenomena have been observed when pressure has been applied to this compound, including a coordination increase from 4- to 6-fold in vanadium atoms, a band-gap collapse of about 1.5 eV, which happens together with a color change, and electrical resistivity dropping at the phase transition [26][29].

Motivated by these interesting phenomena and with the aim to expand the knowledge of the HP behavior of the $CrVO_4$-type family. Here, we report a combined experimental and theoretical study of $CrVO_4$ under high pressure by means of X-ray diffraction (XRD), Raman spectroscopy (RS), optical-absorption measurements, resistivity characterization, and first-principles calculations. All measurements have been carried out up to 10 GPa evidencing a phase transition about 4.5 GPa, which agrees with our theoretical results. Presumably, this phase transition is from the orthorhombic CrVO4-type structure to the wolframite phase that is in good agreement with the north-east rule of the Bastide's diagram [23]. Unit-cell parameters, Raman modes, and band gap energy ($E_{gap}$) are reported under pressure for the low-pressure (LP) and the HP phase. The axial and bulk compressibilities of both phases are also determined. The results are discussed in comparison with the vanadate $InVO_4$. The outcomes of this work may improve the technological applications of the $CrVO_4$ compound and its isomorphic $AXO_4$ compounds.

## EXPERIMENTAL AND CALCULATIONS

Polycrystalline $CrVO_4$ samples were prepared by solid-state reaction from $Cr_2O_3$ and $V_2O_5$ at 723 K following the procedure described by Isasi *et al.* [7]. The crystal structure of $CrVO_4$ was confirmed to be the orthorhombic $CrVO_4$-III (SG: *Cmcm*) structure by means of powder XRD (see Fig. 1). These samples were used for synchrotron powder XRD measurements. For the rest of experiments, we used a polycrystalline sample of $CrVO_4$ which was synthesized by solid-state reaction of appropriate amounts of $(CH_3CO_2)_7Cr_3(OH)_2$ (Sigma-Aldrich) and $NH_4VO_3$ (Panreac). The mixture was ground with an agate mortar. The grounded precursors were heated at 800 °C for 2 h and cooled to room temperature at ambient atmosphere. The obtained compound was characterized by powder XRD and RS, confirming a single phase of $CrVO_4$ with orthorhombic *Cmcm* structure. Minority residuals (~ 1%) of $Cr_2O_3$ and $V_2O_5$ were detected in the synthesized sample. Details of the ambient-pressure crystal structure and Raman spectrum are given in the next section.

Angle-dispersive x-ray diffraction (ADXRD) experiments at room temperature and high pressure were carried out up to 7 GPa at beamline I15 of the Diamond Light Source using a plate-diamond anvil cell (plate DAC) and an x-rays with a wavelength of 0.61486 Å. Samples were loaded in a 200 mm hole of an inconel gasket with diamond-culet sizes of 500 μm. The ruby scale was use for pressure determination [30] and silicone oil was used as PTM [31][32]. Special attention was taken during the experiments to avoid sample bridging between diamonds [33] and the experiments were limited to 7 GPa to limit the influence of non-hydrostaticity in the results. The XRD patterns were collected using a MAR345 image plate placed at 423 mm from the sample. They were integrated and corrected for distortions using FIT2D. The structural analysis was performed using POWDERCELL software.

$CrVO_4$ samples were also studied under compression up to 10 GPa by means of RS, optical absorption and resistivity measurements. Raman experiments were carried out using 16:3:1 methanol-ethanol-water mixture as pressure-transmitting medium (PTM) which guarantee quasi-hydrostatic conditions up to 10 GPa [34]. A plate-DAC with diamond culets of 480 μm was used to generate the pressure. A hardened stainless-steel gasket pre-indented to the thickness of 50 μm with a hole of 200 μm diameter in the center served as the pressure chamber. Pressure was determined using the ruby scale with an accuracy of 0.2 GPa [30].

Optical absorption measurements were done following the same procedure as in previous works [29][35]. A compressed polycrystalline pellet 10 µm thick was placed in a Plate-DAC (the same used for Raman experiment) using the same PTM as Raman experiment and ruby as a pressure gauge. The absorption measurements in the visible-near-infrared range were made using an optical setup specific to determine the optical absorption of wide band-gap semiconductors [35]. The sample-in and sample-out method was used to measure the transmittance and consequently the absorption spectrum [36][37][38].

Resistivity measurements were carried out using an opposed Bridgman-anvil setup as in previous works [29][38]. Hexagonal boron nitride was used as pressure-transmitting medium and the pressure was determined by calibrating the hydraulic pressure versus known phase transitions [39]. The measurements were performed in a compacted powdered sample, 3 mm in diameter and 0.1 mm thickness. Measurements were performed using four ohmic contacts in the Van der Pauw configuration [40].

Calculations were performed within the framework of the density-functional theory (DFT) [41] and the projector-augmented wave (PAW) [42][43] method as implemented in the Vienna Ab initio Simulation Package (VASP) [44][45][46][47]. A plane-wave energy cut-off of 520 eV was used to ensure high precision in our calculations. The exchange-correlation energy was described within the generalized gradient approximation (GGA) in the Perdew-Burke-Ernzerhof for solids (PBEsol) prescription [48]. The GGA+$U$ approximation has been used within the Dudarev's approach ($U_{eff} = U - J^H$) [49] to account for the strong correlation at the electrons in the $d$-orbital. This method has been used with relative success in the study of other vanadates and $ABO_4$ compounds [20][50][51][52][53]. For our calculations, we have used $U$ = 4.5 eV and $J^H$ = 1 eV for Cr atom. Similar values have been used in the study of other Cr based compounds [54][55][56]. Different spin polarizations were considered in the calculations (see Results and Discussion section).

A Monkhorst-Pack grid for Brillouin-zone (BZ) integration [57] has been used with a mesh most suitable for each structure by considering a convergence criterion of 1 meV per formula unit. In the relaxed equilibrium configuration, the forces are less than 1 meV$Å^{-1}$ per atom in each of the Cartesian directions. These highly converged results for the forces are required for the calculations of the dynamical matrix using the direct force constant approach [58]. For the electronic structure, the optimized crystal structures were used with a large set of *k*-points. The phase transition for $CrVO_4$ was obtained by analysing the enthalpy for the phases under study.

## RESULTS AND DISCUSSION

Powder XRD patterns of the as-synthesized $CrVO_4$ sample as well as the patterns under pressure are shown in Fig. 1. The XRD diffractogram of the as-synthesized sample can be well described with the orthorhombic structure (SG: *Cmcm*, No. 63, $Z$ = 4), namely $CrVO_4$-III, having $a$ = 5.568(4) Å, $b$ = 8.208(7) Å and $c$ = 5.977(3) Å as unit-cell parameters. The atomic positions were not refined and were taken from the inorganic crystal structure database (ICSD 36244): Cr: 4*a* (0, 0, 0), V: 4*c* (0, 0.353, 1/4), O1: 8*f* (0, 0.763, 0.99), O2: 8*g* (0.253, 0.49, 1/4). The structure of $CrVO_4$-III can be described as that of $InVO_4$-III [1][26][27][28][59] and is composed of two polyhedral units as building blocks. Vanadium atoms are tetrahedrally coordinated ($VO_4$) and chromium atoms are octahedrally coordinated ($CrO_6$) (see Fig. 2). Along the c-axis, $CrO_6$ octahedra are in contact to each other by edge sharing and form chains connected through $VO_4$ tetrahedral units. On the other side, $VO_4$ tetrahedra are isolated from each other.

All the XRD patterns recorded up to 3.5 GPa correspond to the $CrVO_4$-III phase. Only the usual shift to larger angles of the Bragg peaks is observed due to the unit-cell contraction by compression. In the pattern recorded at 3.5 GPa, it can be observed the appearance of new peaks (labelled with an asterisk in Fig. 1). These peaks coexist with those of the LP phase up to 4.2 GPa. At higher pressure, only the new phase is observed up to 7 GPa (maximum pressure reached in the XRD experiments). As peaks of phase V become predominant as pressure increases, this suggest that a structural phase transition is taking place between 3 and 5 GPa and it is completed at 5.2 GPa.

Due to their structural similarities with $InVO_4$-III and upon crystal-chemistry arguments [23], it can be expected that $CrVO_4$-III would follow the same structural sequence as $InVO_4$-III under compression. The HP phase of $InVO_4$-III is a wolframite-type structure (SG: *P*2/*c*, No. 13, $Z$ = 2) that is isostructural to $InNbO_4$ and $InTaO_4$ [29][35][60]. Our hypothesis is also supported by DFT calculations that show that the most stable structure at ambient conditions is the $CrVO_4$-type (or $CrVO_4$-III) phase, followed by the monoclinic $CrVO_4$-I polymorph. However, when pressure is applied (reduction of sample volume), the most stable structure becomes the wolframite-type structure and not the previously proposed rutile-type phase, which has been obtained after quenching from HP-HT conditions [14][15][16] (see Fig. 3a). Other structures considered as candidates for related compounds [20] have also been included in the calculations (see Fig. 3a), but they are not energetically competitive according to their total energy and enthalpy. The enthalpy pressure evolution of $CrVO_4$-type and wolframite-type structures shows that the $CrVO_4$-type phase is stable up to 3.16 GPa, being the wolframite-type phase the most stable one above this pressure according to thermodynamic arguments (see Fig. 3b). Therefore, our calculations support a phase transition at about 3.16 GPa to a wolframite-type structure, similar to that observed for $InVO_4$-III above 7 GPa [26][27][28]. We would like to add here, that spin polarization was considered in the calculations. The lowest energy state for $CrVO_4$ at ambient pressure and under high pressure is reached only when spin-polarization is taken into account in an antiferromagnetic (AFM) configuration. Where there is an energy difference larger than 2.8 eV/f.u. This is in agreement with ambient-pressure studies of magnetism in $CrVO_4$ [61].

The simulated atomic positions of the wolframite structure of $CrVO_4$ at 7.3 GPa are: Cr: 2*f* (1/2, 0.34541, 1/4), V: 2*e* (0, 0.8321, 1/4), $O_1$: 4*g* (0.22531, 0.11117, 0.07098), $O_2$: 4*g* (0.25845, 0.62195, 0.09325) and the unit-cell parameters are given in Table 1. The theoretical parameters obtained by DFT calculations for the wolframite-type structure were used to refine the XRD patterns of the HP phase giving a satisfactory fit (see Table 1 and Figure 4). Rietveld refinement was carried out following the steps described in Ref. [26]. In Table 1, we provide the refined unit-cell parameters and volume of the experimental wolframite phase at 7 GPa. We also include the bulk modulus and its pressure derivative (at ambient pressure) as obtained from a Birch-Murnaghan (BM) equation of state (EoS) fit. For comparison, we included the parameters of other HP wolframites, such as $InVO_4$ and $FeVO_4$, that we will discuss later in the text. These data is given at the pressure reported in the literature [20][26][27][28][62]. The wolframite crystal structure of $CrVO_4$ can be seen in Fig. 2. The structure corresponds to the space group *P2/c* (No. 13, $Z$ = 2). The structural phase transition is accompanied by large volume collapse of 14%. Analyzing the structural parameters, an increase in vanadium atoms coordination from 4 (tetrahedral) to 6 (octahedral) with no notorious change in the octahedral coordination of chromium was observed (see Fig. 2). However, in the new configuration, the $VO_6$ octahedra units are more distorted compared to $CrO_6$ octahedral units. These structural modifications agree very well with the observed structural modification in $InVO_4$-III under compression [26][27][28].

The pressure dependence of the unit-cell parameters and volume for the LP and HP phases of $CrVO_4$-III are shown in the top and bottom of Fig. 5, respectively. In the LP phase, a clear anisotropy in the axial compressibility is observed being the highest variation under pressure in the *b*-axis. As discussed in Ref. [26], this fact is related with the absence of $VO_4$ tetrahedral units along the *b*-axis direction and the enhancement of the $CrO_6$ octahedral distortion under pressure compared to $VO_4$ tetrahedral. The distortion index parameter of octahedral units $CrO_6$ vary 35 % between 0.7 – 4.2 GPa. On the contrary, the tetrahedral units $VO_4$ vary 9 % in the same pressure range (here, it is used the Baur distortion index [63] as implemented in VESTA software used for crystal structure visualization [64]). The distortion occurs due to the major compressibility of Cr-O bonds that are aligned along the *b*-axis, as compared with equatorial bonds (plane containing the *a*- and *c*-axis) of the octahedral units. Consequently, the crystal structure is more compressible along *b*-axis. The reduction of Cr-O bonds along the *b*-axis in the pressure range 0.7 – 4.2 GPa is about 4 %, whereas the reduction of the equatorial bonds is 0.8 %. In summary, $CrO_6$ octahedral units become flattened under compression.

The pressure dependence of the unit-cell parameters of the LP phase can be well described by the linear equations:

$a$ (Å) = 5.559(4) – 7.0(2) × $10^{-3}$ P

$b$ (Å) = 8.218(7) – 7.2(3) × $10^{-2}$ P

$c$ (Å) = 5.976(3) – 2.1(2) × $10^{-2}$ P

with P in GPa. The axial compressibilities at ambient pressure are: $\kappa_a = 1.3 \times 10^{-3}$ $GPa^{-1}$, $\kappa_b = 8.8 \times 10^{-3}$ $GPa^{-1}$, $\kappa_c = 3.5 \times 10^{-3}$ $GPa^{-1}$.

The pressure dependence of the unit-cell parameters of the HP phase can be well described by the linear equations:

$a$ (Å) = 4.378(2) – 1.3(4) × $10^{-3}$ P

$b$ (Å) = 5.594(12) – 2.7(2) × $10^{-2}$ P

$c$ (Å) = 4.723(10) – 3.2(9) × $10^{-3}$ P

$\beta$ (°) = 89.32(6) + 0.26(2) P – 1.8(2) × $10^{-2}$ $P^2$

with P in GPa.

Notice that the $\beta$ angle is close to 90º and change little with pressure (see Fig. 5), so the change with pressure of the unit-cell can be described in first approximation by the axial compressibilities, which are $\kappa_a = 0.3 \times 10^{-3}$ $GPa^{-1}$, $\kappa_b = 4.8 \times 10^{-3}$ $GPa^{-1}$, $\kappa_c = 0.6 \times 10^{-3}$ $GPa^{-1}$. Again, it can be observed that *b*-axis is the most compressible while the *a*-axis is almost incompressible. The unit-cell parameters follow the same trend as observed in $InVO_4$ [26]. In Fig. 5, it is also included the theoretical pressure dependence of the unit-cell parameters. It can be seen the good agreement between experiments and theory.

In Fig. 5 (bottom), we show the pressure dependence of the unit-cell volume of the LP and HP phases of $CrVO_4$-III. The volume of the HP phase has been doubled for a better comparison with that of the LP phase. In both phases, a second order BM EoS has been employed to fit the experimental data. For the LP phase, we have obtained a zero-pressure experimental (*theoretical*) volume of $V_0$ = 273.0(5) / *(279.8)* Å$^3$ and bulk modulus of $B_0$ = 63(5) / *(93)* GPa, being $B_0'$= 4 the bulk modulus pressure derivative. The bulk modulus is similar to that of the orthorhombic phase of $InVO_4$ ($B_0$ = 69(1)

GPa), despite the zero-pressure volume of $InVO_4$ is 23 % bigger than that of orthorhombic $CrVO_4$. This result can be understood if we consider that the $Cr^{3+}$ atom has a smaller Shannon ionic radius than the $In^{3+}$ atom and this leads to a reduction of the octahedral volume around 19 % and of the tetrahedral unit around 5 %. These differences make that octahedra in $CrVO_4$-III are more distorted than those of $InVO_4$-III despite both compounds have similar distorted tetrahedra.

The parameters of the BM EoS fit used for the HP phase can be seen in Table 1. There is a good agreement in the unit-cell parameters and unit-cell volume between theory and experiment. However, our calculations have a tendency to overestimate the bulk modulus. The increment of bulk modulus from the orthorhombic phase to the wolframite phase is consistent with the observed large volume collapse and an increase of the HP phase density due to the reduction of empty space between polyhedra due to the increase of vanadium coordination. Interestingly, $InVO_4$, $FeVO_4$, and $CrVO_4$ have an isomorphic wolframite-type structure with similar bulk moduli. Moreover, we can reasonably speculate that a wolframite-type polymorph could also exist at relative LP in $TlVO_4$.

To further investigate the behavior of $CrVO_4$ under pressure, Raman spectroscopy experiments were carried out and selected spectra at different pressures are shown in Fig. 6. These experiments will be analyzed with the help of DFT calculations, which give not only Raman frequencies and mode assignment, but also the phonon dispersion and density of states (PDOS). They are displayed in Fig. 7. Our phonon calculations show that the ambient-pressure and HP phases of CrVO4 are both dynamically stable. The Raman spectrum of $CrVO_4$-III at room pressure resembles to those previously reported [65][66][67]. According to the literature and the group theory analysis, $CrVO_4$ with orthorhombic *Cmcm* structure has 15 Raman active modes at Γ point: $\Gamma = 5A_g + 4B_{1g} + 2B_{2g} + 4B_{3g}$. We could identify the 15 modes and they agree well with our theoretical calculations (see Fig. 8. and Table 2). The mode assignment was based on our lattice dynamic calculations. Several features can be identified in the Raman spectrum of $CrVO_4$-III at room pressure. For instance, the most intense peaks are the modes at the highest frequency and there is a phonon gap between 500 and 650 $cm^{-1}$. As in most $ABO_4$ compounds, the Raman-active modes of $CrVO_4$-III can be classified as internal or external modes of the tetrahedral $VO_4$ units. The high frequency modes (924-930 $cm^{-1}$) have been assigned to the symmetric bending ($\nu_1$) and asymmetric stretching ($\nu_3$) of V-O bonds. On the other side, the low frequency modes (168-248 $cm^{-1}$) are due to pure translation (T) of $VO_4$ tetrahedra units. The intermediate modes (277-378 $cm^{-1}$) correspond to pure rotation (R) and symmetric bending ($\nu_2$) [65][66][67]. The modes (381-502 $cm^{-1}$) are related to symmetric ($\nu_2$) and asymmetric ($\nu_4$) bending modes. It must be added that the three high-frequency modes above 700-750 $cm^{-1}$ can be assigned to the symmetric ($\nu_1$) and two asymmetric ($\nu_3$) stretching modes of the deriving from the isolated $VO_4$ tetrahedra, as already found in zircon-type orthovanadates, such as $YVO_4$ [68], $ScVO_4$ [69], $CeVO_4$ [50], $TbVO_4$ [70] and $NdVO_4$ [71] among others. In fact, internal and external modes of $CrVO_4$-type compounds have been fully assigned in $InVO_4$-III [26]. It must also be stressed that the Raman mode of highest intensity in CrVO4-type compounds $CrVO_4$-III and $InVO_4$-III is the second highest frequency mode while in zircon-type compounds is the highest frequency mode. In both cases, it corresponds to the symmetric ($\nu_1$) stretching mode.

In the Raman spectra, $CrVO_4$-III is the only phase present up to 1.6 GPa (zone I in Fig. 9). At 1.8 GPa, the first signature of the HP phase is observed (zone II in Fig. 9). The most intense peak of the HP phase appears around 808 $cm^{-1}$ as a shoulder of the $\omega_{13}$ (765 $cm^{-1}$) mode (see Fig. 6). Although, this peak is very weak and cannot be appreciably seen up to 2.8 GPa, it can be followed by peak fit analysis (see Fig. 9). Here, we should mention that the phase transition pressure in Raman experiments is slightly lower than in XRD experiments. A smaller transition pressure in Raman experiments than in XRD experiments is always expected due to the more local character of Raman scattering than of XRD [72]. However, we cannot rule out sample bridging between diamonds in our RS experiments. This effect would reduce the transition pressure due to non-hydrostatic stresses and would also explain the considerable broadening of peaks observed in the HP phase as well as the long coexistence pressure range of both phases (zone II, III and IV in Fig. 9). Additional features of the HP phase are seen between 2.8 and 3.4 GPa in the 375-474 $cm^{-1}$ range. Then, the rest of the HP phase features appears clearly in the pressure range 3.8-5.5 GPa. At this pressure, the most prominent peak of the HP phase becomes the predominant one (zone III in Fig. 9). It must be noted that the two phases coexist up to 8.5 GPa (zone IV in Fig. 9) and only the HP phase is present at higher pressures up to 11 GPa (zone V in Fig. 9). On pressure release, the ambient pressure spectrum is recovered but showing small presence of the HP phase and contribution of chromium oxide (Fig. 6). This result manifests the partial reversibility to $CrVO_4$-III upon decompression. It must be stressed that the existence of phase mixture was also observed in $InVO_4$-III on pressure release [26].

The evolution of Raman modes under pressure can be described using a linear fit in good agreement with our theoretical calculations (see Fig. 9). Upon compression, almost all Raman modes shift to higher frequencies, but the mode labelled $\omega_2$ shows a gradual softening (see Table 2). This mode softening was not observed in its partner $InVO_4$-III [26]; however, soft modes have been observed between the external modes of other orthovanadates [68][69][50][70][71]. In particular, $ScVO_4$ [69] shows two soft modes, one of them being the second lowest-frequency mode, like in $CrVO_4$.

On the other hand, $\omega_{12}$ is the mode more sensitive to pressure; i.e., the Raman vibration to which more energy is transferred during compression. This is in contrast to what happens in zircon-type orthovanadates, where the stretching modes show the largest pressure coefficients. Another curious thing is that in $CrVO_4$-type compounds $CrVO_4$-III and $InVO_4$-III the symmetric ($\nu_1$) stretching mode has a very small pressure coefficient (below 1.5 $cm^{-1}$/GPa), while in zircon- and scheelite-type orthovanadates it shows one of the largest pressure coefficients [68][69][50][70][71]. Finally, we must note that the $\omega_6$ mode presents a huge dispersion due to the difficulty to follow it under compression since it is a mode that appears as a shoulder in the left side of the $\omega_7$ mode.

As it was commented previously, the major changes in the Raman spectra occur in the pressure range 3.5-5.5 GPa (zone III in Fig. 9). At 5.5 GPa, most of the material shows the characteristic Raman features of the HP phase, with the peak around 800 $cm^{-1}$ becoming the predominant one. Thus, it reasonable to consider that the phase transition is taking place in this pressure range. This is consistent with the pressure range observed in the XRD data and with the optical and resistivity measurements that we will discuss later.

For the HP phase wolframite-type having *P2/c* symmetry, group theory predicts 18 Raman active modes: $\Gamma = 8\ A_g + 10\ B_g$. The HP phase modes were analyzed, although, the overlapping of the peaks with the LP phase and peak broadening make difficult to distinguish all the Raman features. Thus, we are unable to follow some modes under pressure (see Table 3). However, at the highest pressure with the help of our Raman calculations, we are able to identify and assign all the theoretically predicted Raman modes. The HP phase Raman spectrum resembles that in $InVO_4$-V [26] showing the most intense peak at the highest frequency. This is different than for other wolframites-type structures such as $InNbO_4$ or $InTaO_4$, where the most intense peaks are in low frequency range [26][35][60], but very similar to what happen in wolframite-type tungstates [73]. Since high-frequency modes are associated to internal vibrations of the $VO_6$ units, the observed mode intensities suggest that the polarizability of the vanadate (and tungstate) molecules are more sensitive to the electric field of light than in niobates and tantalates. Comparing the pressure evolution of the Raman-active modes in region III and V (see Fig. 9), one can observe the softening of the mode at the highest frequency. This result implies that the $VO_4$ units change its coordination, which is associated to a phase transition, as observed by XRD experiments. This mode softening has also been observed in the wolframite phase of $InVO_4$, which is coherent with the assignment of wolframite-type structure done to $CrVO_4$ at HPs.

In order to study the effects of pressure in the electronic structure of $CrVO_4$ under compression, theoretical calculations and optical absorption measurements were carried out. Fig. 10 (top) displays the optical-absorption coefficient ($\alpha$) as a function of energy of the LP phase at several pressures. A step-like absorption edge can be observed at ambient condition. This feature corresponds to the fundamental absorption between the maximum valence band (VBM) and the minimum of the conduction band (CBM), together with a low-energy absorption tail. This tail is usually described with an Urbach equation [74], which is associated with presence of defects in the lattice crystal structure. These defects may introduce localized levels in the band gap giving rise to a possible under estimations of the fundamental band gap energy [29][35]. This Urbach tail has been observed for many ternary oxides as well as for $InVO_4$-III compound [29][35][74][75][76][77]. The high energy region of the absorption edge follows a linear behavior when it is plotted as $(\alpha E)^2$ versus photon energy (E), which is analyzed by using a Tauc plot [78]. This is consistent with a direct band gap nature and agrees with the values reported in the literature and given by our theoretical calculations (see Fig. 10, bottom and table 4) [10][67]. The energy band gap is estimated at 2.62(5) eV and is consistent with our calculations and with the value reported by Yan *et al* [10]. However, our experimental energy band gap is slightly higher but consistent with Bera *et al* work reporting an energy band gap of 2.44 eV [67] and up to 1 eV different from the energy band gap estimated indirectly from the calculated activation energy measured by Gupta et al [79]. This discrepancy may be due to the large uncertainty associated to the last method.

Band structure calculations of $CrVO_4$-III shows two indirect transitions at T-Γ and R-Γ *k*-directions at lower energies than the direct transition at Γ-Γ *k*-direction (see Fig. 11, a). However, an indirect transition needs the participation of phonon to conserve momentum and giving the proximity of the indirect and direct transitions, the oscillator strength of the direct transition must be predominant. Thus, by optical absorption measurements, we only observe the direct transition as it is more likely to occur. Based on our calculations, the energy gap from indirect transitions are 2.42 eV at T-Γ *k*-direction and 2.38 eV at R-Γ *k*-direction. Our estimated energy band gap value is consistent with the value obtained from calculations at Γ-Γ *k*-direction. According to density of states (DOS) calculations, VBM of $CrVO_4$-III is dominated by a mixture of oxygen 2p states and chromium 4s states. On the other hand, CBM is mainly dominated by vanadium 3d states with a minor contribution of Cr and O states (see Fig. 12).

Upon compression, the fundamental edge absorption of $CrVO_4$-III blue-shifts up to 4.7 GPa (see Fig. 10, inset). Above this pressure, a sudden drop in energy band gap is experimentally observed in the absorption edge (see Fig. 10 and 13 for comparison), which confirms the existence of a phase transition. Fig. 13 shows the optical-absorption edge of the HP phase at several pressures together with the $E_{gap}$ pressure evolution (inset). The evolution of $E_{gap}$ vs pressure follows a linear behavior with a rate of 1.9(3) meV/GPa for the $CrVO_4$-III phase. In the pressure range 4.0-5.5 GPa (colored area in Fig. 13, inset), the $E_{gap}$ drops about 1.1 eV which is associated to a band gap collapse due to the phase transition observed previously by XRD and Raman spectroscopy measurements. Again, using Tauc´s analysis, the HP phase energy gap was estimated at 1.53(5) eV showing a direct transition character (see Fig 13, bottom). As discussed for the LP phase, our estimated energy band gap value for the HP phase corresponds to the Γ-Γ *k*-direction, although, other lower energy indirect transitions are predicted by our calculations about 0.87 eV at C - Γ and 0.85 eV at A0 - Γ *k*-directions (see Fig. 11, b). DOS calculation shows that VBM is dominated by Cr and O states and the CBM by V states but, with lower contribution from Cr and O states, in this case (see Fig. 12, bottom). After the $E_{gap}$ collapse, the band gap redshifts following a linear behavior. Similar band gap collapse was observed for $InVO_4$-III at the phase transition and in many other ternary oxides [26][29][35]. However, in the case of $InVO_4$-V, the $E_{gap}$ blue-shifts under compression. An additional interesting fact that can be concluded from electronic band-structure calculations is that in $CrVO_4$ under pressure, there is an enhancement of the hybridization of O 2p orbitals with Cr 4s orbitals, which is the responsible to favor the stability of the high-pressure phase and band-gap closing at the phase transition.

To obtain more information from the optical absorption-measurements, the Urbach energy (Eu) was analyzed under pressure and the data are shown in Fig. 14 and Table 5. Eu was estimated following the protocol described in our previous works [29][80]. Eu increases with the pressure and around 4.5 GPa drops 24 meV. After this drop, Eu decreases with pressure. The same as in the optical-absorption measurements, this drop is related to the phase transition. The Eu of $CrVO_4$-III is estimated to be 59(1) meV at room pressure (see Table 5) and its evolution under pressure can be interpreted as due to the appearance of defects (for instance dislocations) in the pressure range 0.7-4.2 GPa. On the contrary, in the HP phase the structure under pressure the Urbach energy decreases with pressure in the pressure range 4.2 to 9.5 GPa where the HP phase exist. The HP phase Urbach energy is 52 meV and decreases up to 42 meV in the above-mentioned pressure range, suggesting that the degree of disorder is reduced under compression in the HP phase.

In Fig. 15, we plot the electrical resistivity measurements up to 10 GPa. $CrVO_4$-III is a p-type semiconductor with a conduction dominated by holes and in extrinsic regime, due to a small polaron hopping. The conduction mechanism has been associated with the presence of impurities and defects in the crystal structure, specifically with cation-deficient centers [79][81]. At ambient conditions, $CrVO_4$-III shows a resistivity of the order of 1 k$\Omega$·cm that is three order of magnitude smaller than isostructural $InVO_4$-III with a resistivity of the order of 1 M$\Omega$·cm [29]. The electrical resistivity decreases under compression, in the LP phase (from 1 bar to 3.5 GPa). This can be caused by the enhancement of polaron hopping or due to the formation of defects that introduce acceptors levels and increase the carrier concentration. Consequently, the activation energy decreases with pressure at a rate of -2.5(2) meV/GPa. Even though, the bang-gap increases in energy under pressure, the changes on VBM favors the decrease of the energy difference between the acceptor level and the VBM (most probably via a change of the hole effective mass [82]). In the pressure range from 3.5 to 4.6 GPa (blue-dashed area in Fig. 15), a drastic drop of the resistivity is observed. The pressure range where the electrical resistivity drop occurs, agrees with that of the band-gap collapse and the structural phase transition from $CrVO_4$-III to $CrVO_4$-V observed by XRD and Raman scattering. The measured-points in the blue-dashed area suggest a phase coexistence during the transition. In the HP wolframite phase, the resistivity continues dropping, with a decrease of the activation energy at a rate of -4.9(3) meV/GPa until the maximum pressure is reached in the experiment. The decrease of the resistivity could be related to the creation of oxygen vacancies along the transition [83], which will favor the increase of the free-carrier concentration, reducing this way the electrical resistivity.

## CONCLUSIONS

In this work, we have reported evidence of a pressure-induced phase transition in orthorhombic $CrVO_4$ at room temperature. The phase transition to a new monoclinic polymorph isostructural to wolframite takes place near 4 GPa. We have also reported results from the influence of pressure in the phonons, optical, and electronic properties of both the low and high-pressure phases of $CrVO_4$. In particular, a drastic decrease is found in the band-gap and resistivity. The Raman spectrum qualitative change at the transition, with the changes being associated to the symmetry decrease and V coordination number increase that occur at the phase transition. The room-temperature equation of state has been determined for the two phases of $CrVO_4$. Experimental results are consistent with results from density-functional calculations, which are crucial for the interpretation of experiments. Finally, the behavior of $CrVO_4$ under high-pressure is compared with related oxides being general conclusion for the family of $CrVO_4$-type compounds proposed.

## ACKNOWLEDGEMENTS

This work was supported by the Spanish Ministry of Science, Innovation and Universities under grants MAT2016-75586-C4-1/2-P, FIS2017-83295-P and RED2018-102612-T (MALTA Consolider-Team network) and by Generalitat Valenciana under grant Prometeo/2018/123 (EFIMAT). P. B. and A. V. acknowledge the Kempe Foundation and the Knut och Alice Wallenberg Foundation for their financial support. J. A. S. also acknowledges Ramón y Cajal program for funding support through RYC-2015-17482. The x-ray diffraction measurements were carried out with the support of the Diamond Light Source at the I15 beamline under proposal No. 683. The authors thank A. Kleppe for technical support during the experiments. S. L.-M. thanks CONACYT of Mexico for financial support through the program "Cátedras para jóvenes Investigadores". Also, S. L.-M. gratefully acknowledges the computing time granted by LANCAD and CONACYT on the supercomputer Miztli at LSVP DGTIC UNAM. Besides, some of the computing for this project was performed with the resources of the IPICYT Supercomputing National Center for Education & Research, grant TKII-R2020-SLM1.

## FIGURES CAPTIONS

**Fig. 1.** Powder XRD patterns of $CrVO_4$ at selected pressures. Ticks indicate the position of Bragg peaks of phases III and V.

**Fig. 2.** Crystalline structure of orthorhombic (left) and wolframite-type (right) $CrVO_4$. Vanadium atoms and corresponding polyhedral coordination are shown in red. Chromium atoms and corresponding polyhedral are shown in blue. Oxygen atoms are represented in orange. (The reader is referred to the online version for color code).

**Fig. 3.** (a) Energy-volume curves of proposed $CrVO_4$ structures calculated with PBEsol+*U* approximation. (b) Pressure dependence of enthalpy for the $CrVO_4$- and wolframite-type phases of $CrVO_4$.

**Fig. 4.** Rietveld refinement plots for orthorhombic (phase III) and wolframite-type (phase V) of $CrVO_4$. Dots: experimental data. Solid lines: refinement. Dashed lines: residuals. Ticks indicated calculated positions of Bragg reflections. *hlk* indices are also shown.

**Fig. 5.** (Top) Pressure dependence of unit-cell parameters of orthorhombic (solid symbols) and wolframite-type (empty symbols) of $CrVO_4$. The inset shows the β angle evolution with pressure. (Bottom) unit-cell volume evolution with pressure for orthorhombic (solid symbols) and wolframite-type phase (empty symbols). Dashed lines represent the fits and equation of state in the case of volume. Solid lines represent the theoretical calculations.

**Fig. 6.** Selected high pressure Raman spectra on $CrVO_4$. Pressure steps indicated in the right side of the plot. The asterisk points out a peak coming from residuals of $V_2O_5$. In the recovered spectrum is pointed out by arrows peaks from the HP phase and by & contribution of residuals chromium oxides.

**Fig. 7.** Phonon dispersion (left) and phonon density of states (right) for both phases. Calculations were done at ≈0 (top) and 7.3 GPa (bottom) for CrVO4-type and wolframite, respectively.

**Fig. 8.** Raman spectra for orthorhombic (phase III) and wolframite-type (phase V) of $CrVO_4$. Calculated Raman frequency modes are shown under data by tick symbols.

**Fig. 9.** Pressure dependence of phonon wavenumber extracted by peak fitting the Raman spectra. The plot is divided in four regions: (I) existence of the LP phase $CrVO_4$-type, (II) coexistence of both phases being $CrVO_4$-III, the predominant phase, (III) region where major changes occur, (IV) coexistence of both phases being $CrVO_4$-V, the predominant phase and (V) existence of the HP phase wolframite. Solid and dashed lines represent the theoretical calculation evolution.

**Fig. 10.** (top) Absorption spectra measured at different pressures for the LP phase of $CrVO_4$. The inset shows a zoom of the highest values of absorption coefficient.

(bottom) Tauc plot used to determine *E*g. The dashed line shows the extrapolation of the linear region to the abscissa.

**Fig. 11.** Band structure for (a) $CrVO_4$-type and (b) wolframite phase.

**Fig. 12.** Density of states (DOS) for (a) $CrVO_4$ and (b) wolframite phase.

**Fig. 13.** (top) Absorption spectra measured at different pressures for the HP phase of $CrVO_4$. The inset shows *E*g versus pressure for the LP (square symbols) and HP (triangle symbols) phases. The band-gap collapse is indicated. The symbols are the experimental results, and the solid(dashed) line is the linear fit to them for LP(HP) phase respectively. Solid and dashed lines are the theoretical calculations for the respective phases. (bottom) Tauc plot used to determine *E*g. The dashed line shows the extrapolation of the linear region to the abscissa.

**Fig. 14.** Urbach energy data of $CrVO_4$ measured at different pressures obtained from fitting the absorption coefficient data. Solid symbols for the LP phase and empty symbols for the HP phase. Solid line represents the linear fit to the data.

**Fig. 15.** Resistivity as a function of pressure measured for $CrVO_4$.

**FIGURES**

**Fig. 1.**

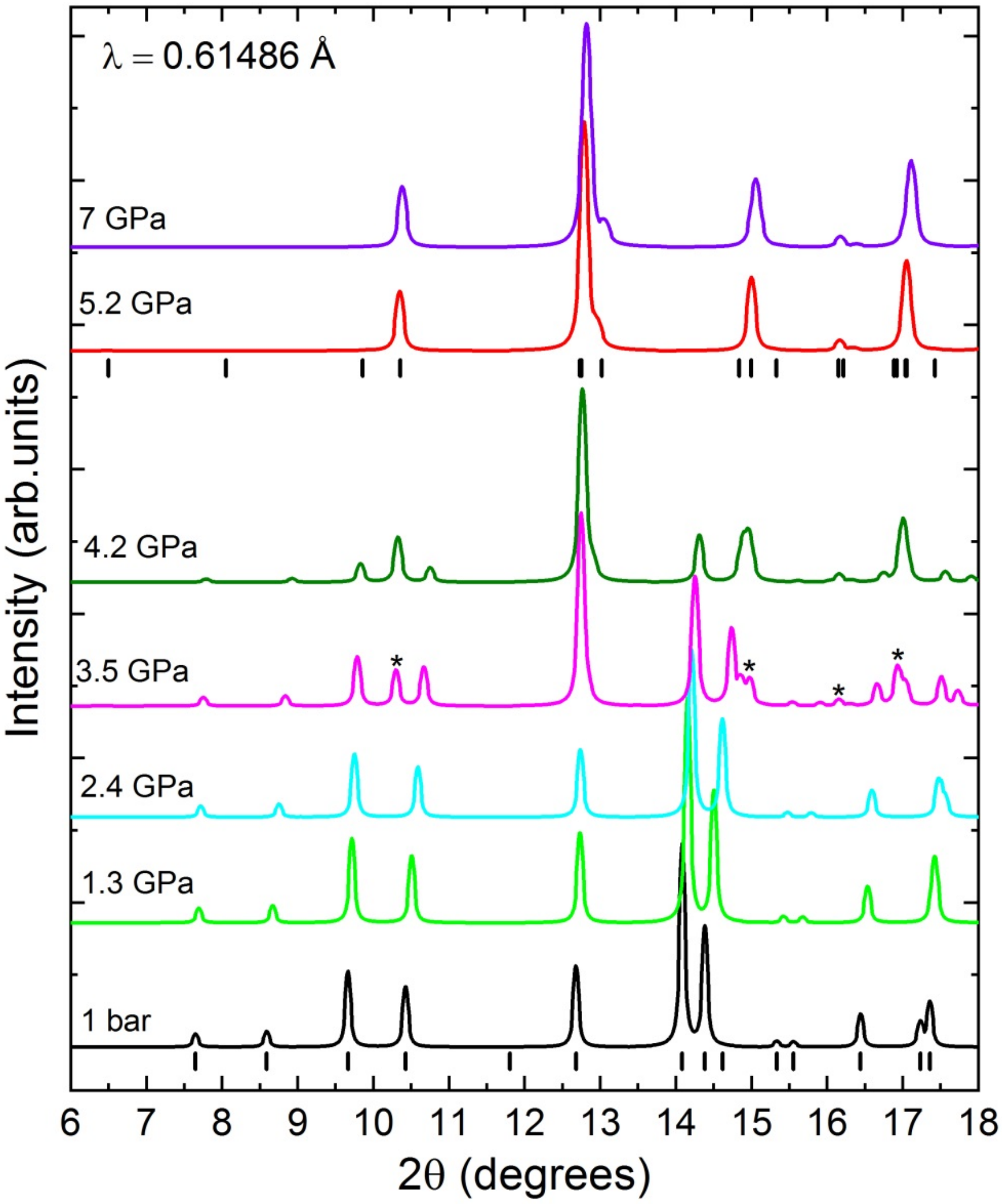

**Fig. 2.**

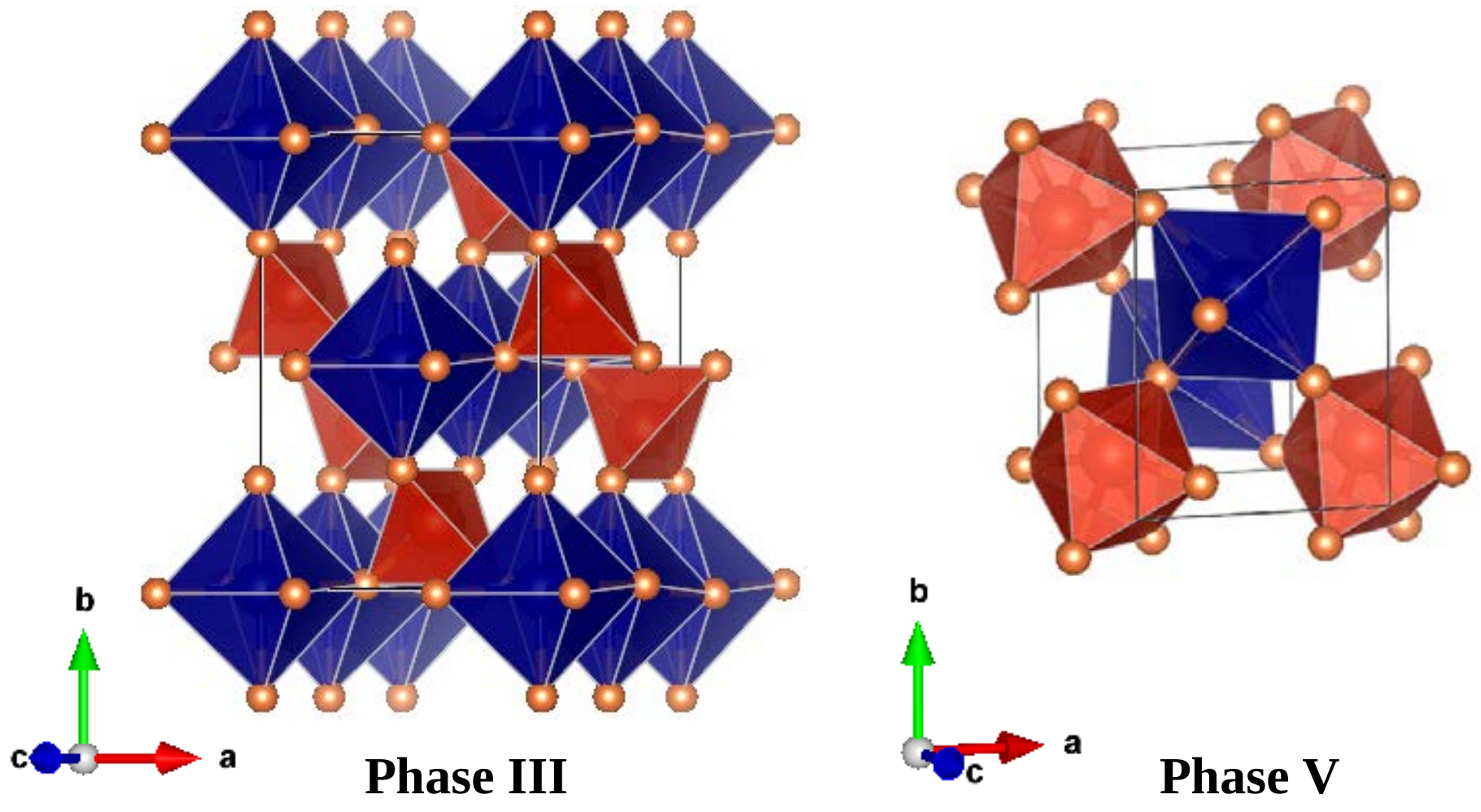

**Fig. 3**.

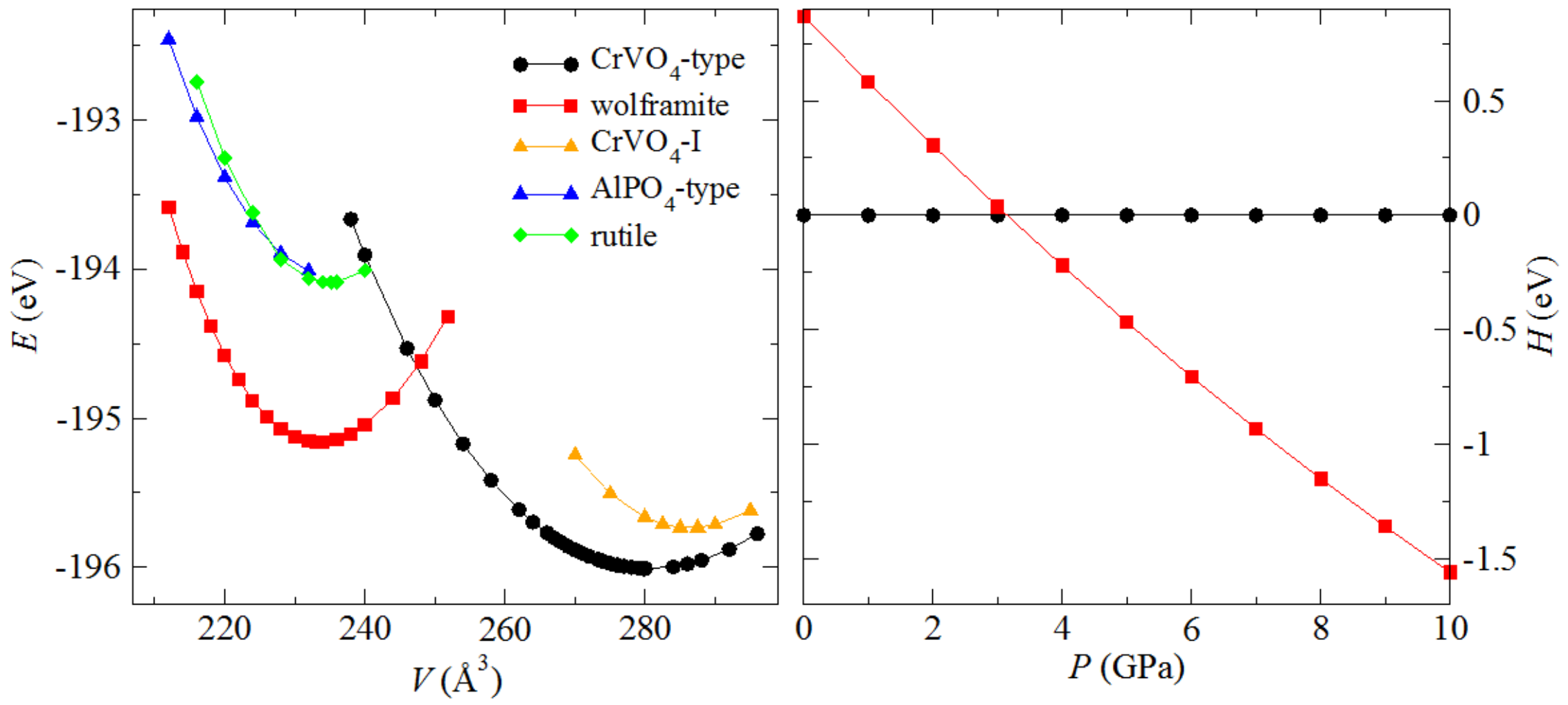

**Fig. 4.**

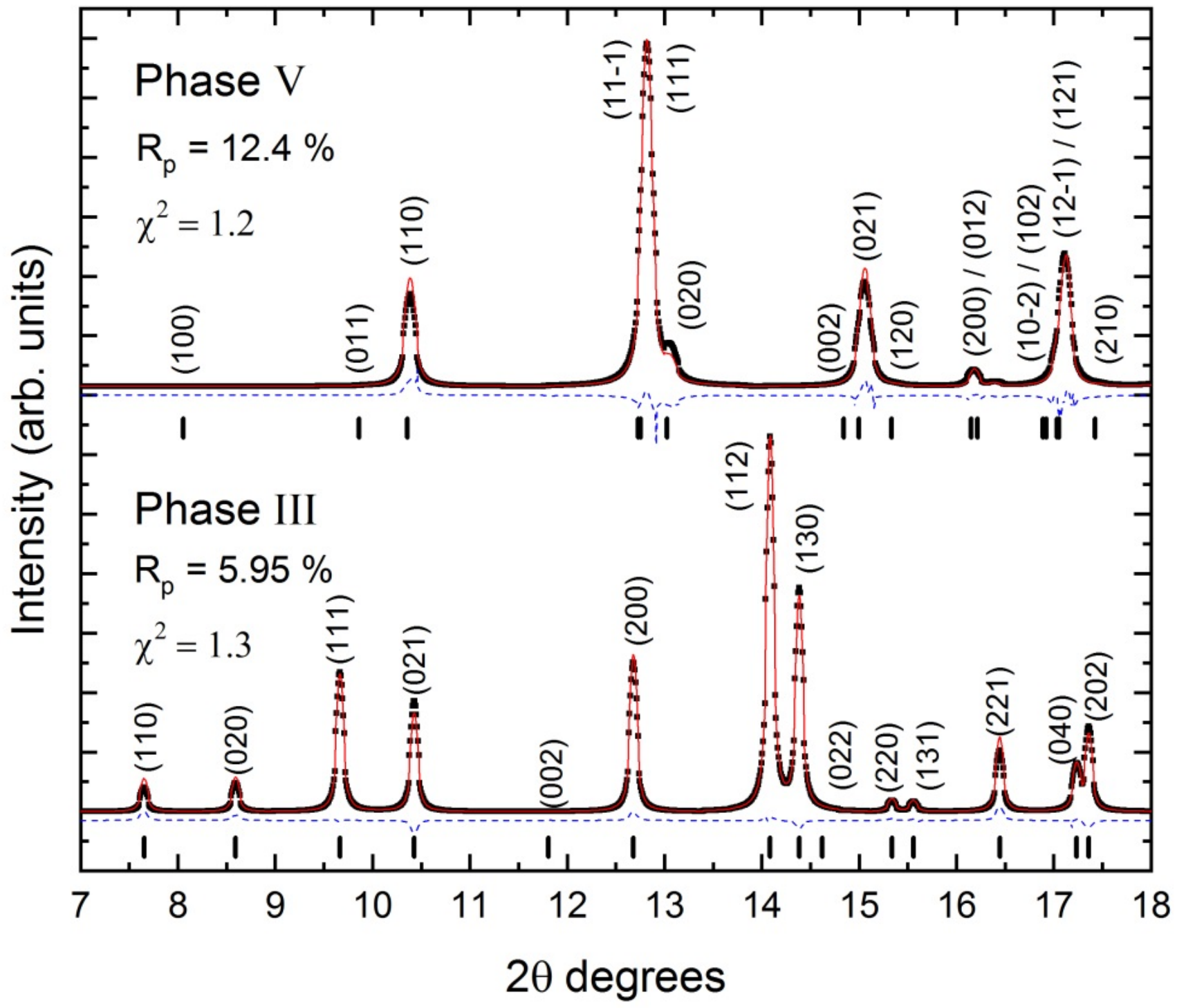

Phase V
$R_p$ = 12.4 %
$\chi^2$ = 1.2
(100)
(011)
(110)
(11-1)
(111)
(020)
(002)
(021)
(120)
(200) / (012)
(10-2) / (102)
(12-1) / (121)
(210)
Phase III
$R_p$ = 5.95 %
$\chi^2$ = 1.3
(110)
(020)
(111)
(021)
(002)
(200)
(112)
(130)
(022)
(220)
(131)
(221)
(040)
(202)
Intensity (arb. units)
7 8 9 10 11 12 13 14 15 16 17 18
2θ degrees

**Fig. 5.**

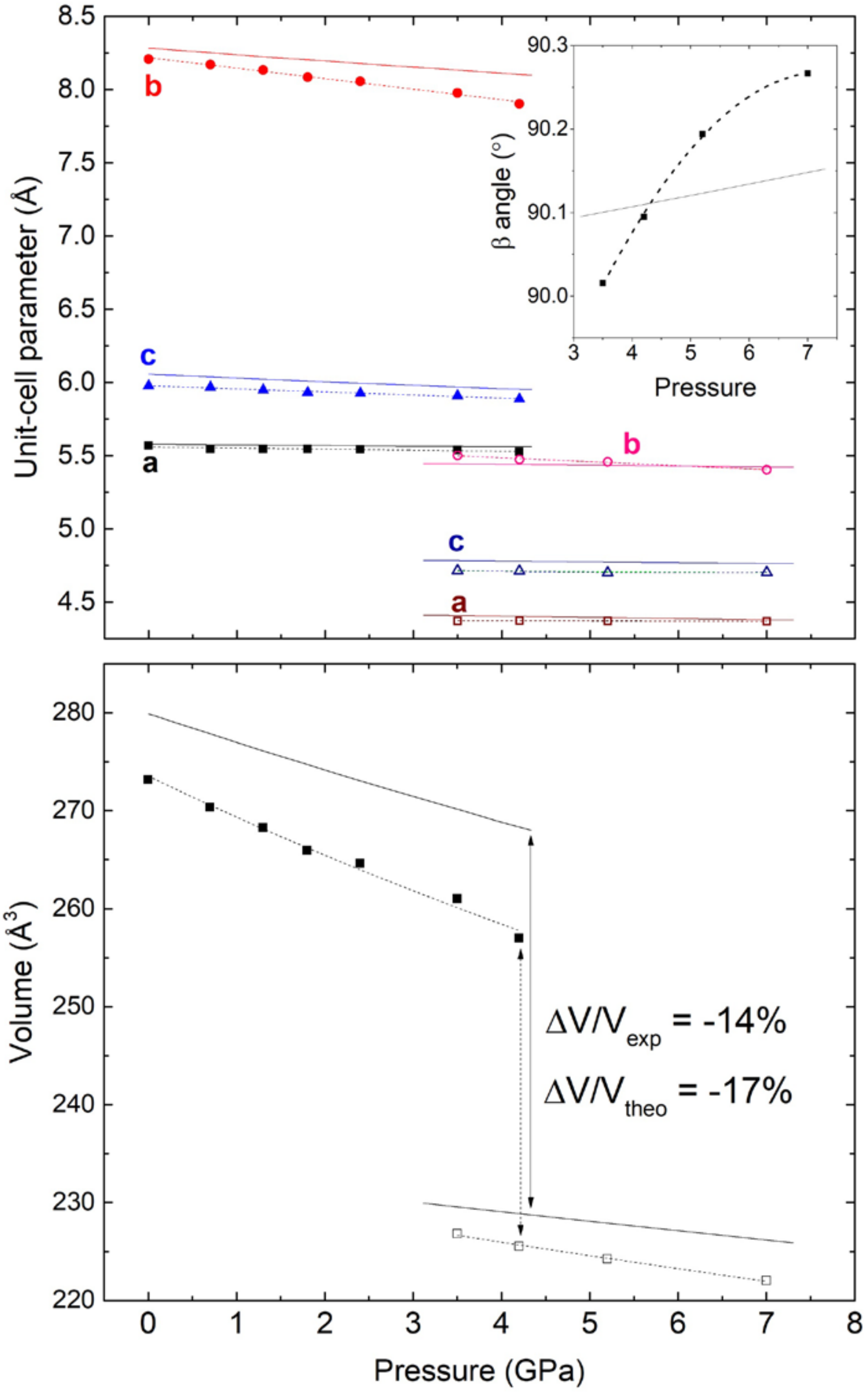

**Fig. 6.**

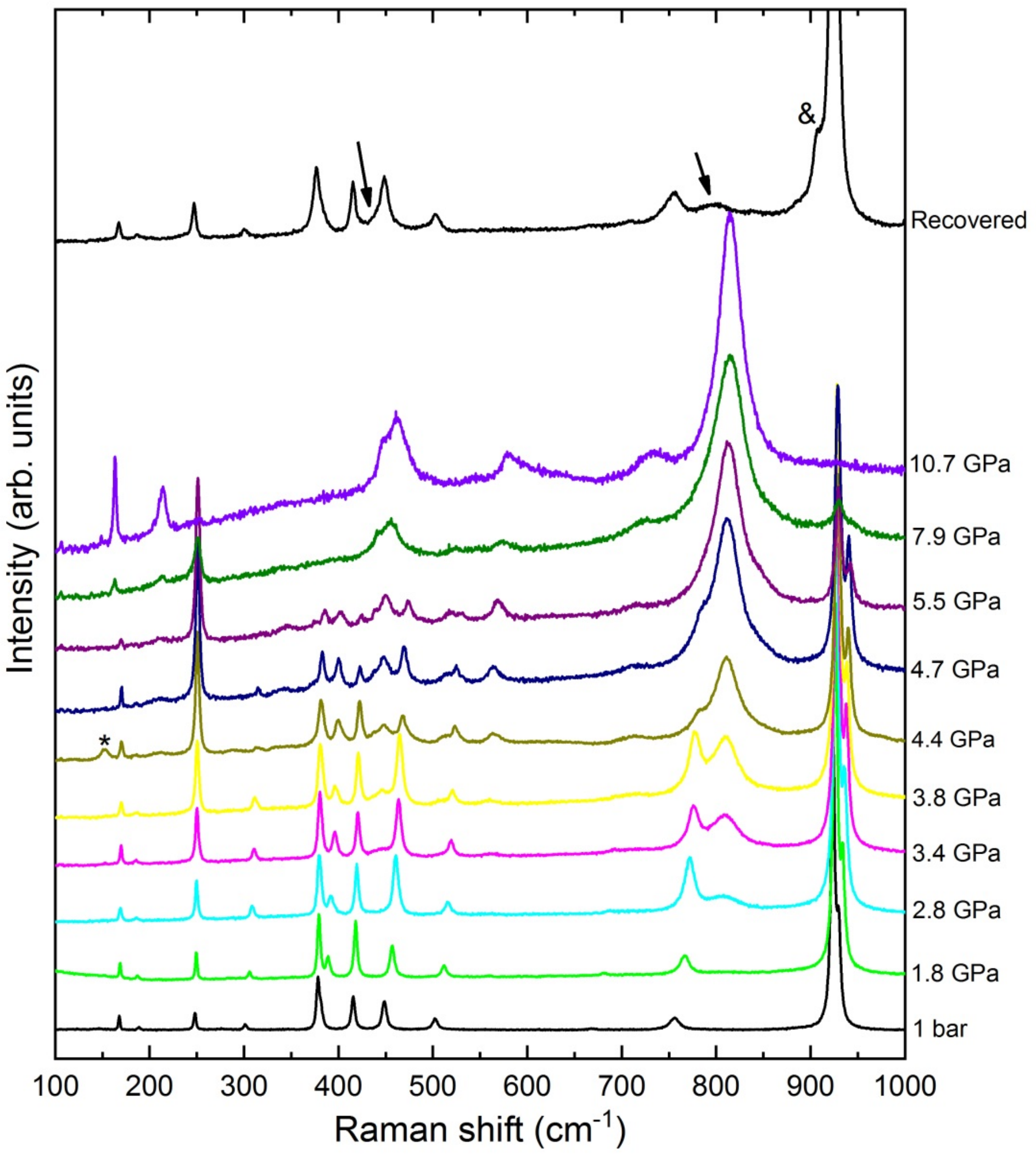

**Fig. 7.**

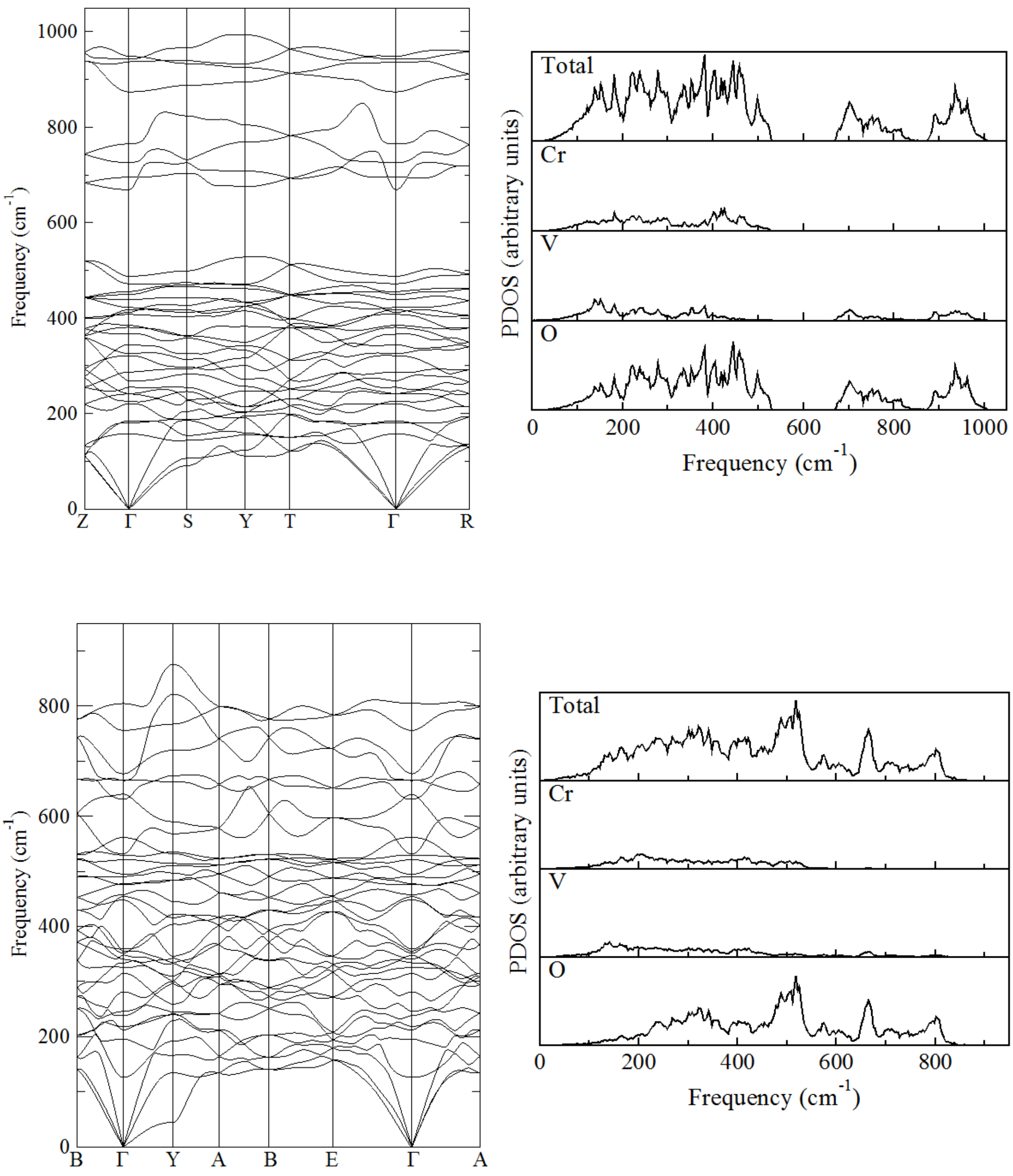

Frequency (cm-1)
Z Γ S Y T Γ R
PDOS (arbitrary units)
Total
Cr
V
O
Frequency (cm-1)
Frequency (cm-1)
B Γ Y A B E Γ A
PDOS (arbitrary units)
Total
Cr
V
O
Frequency (cm-1)

**Fig. 8.**

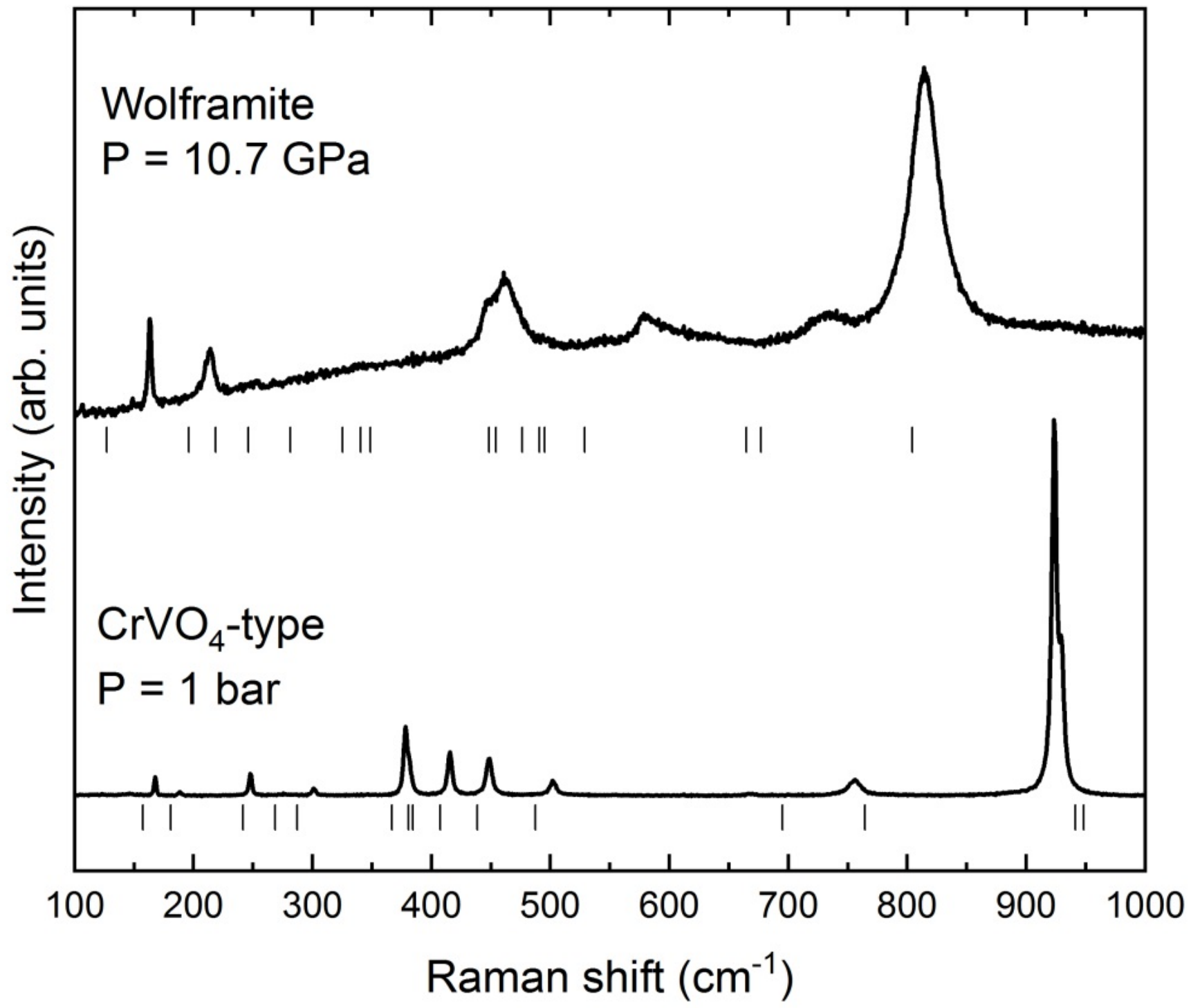
Wolframite
P = 10.7 GPa
CrVO4-type
P = 1 bar
Intensity (arb. units)
100 200 300 400 500 600 700 800 900 1000
Raman shift (cm-1)

**Fig. 9.**

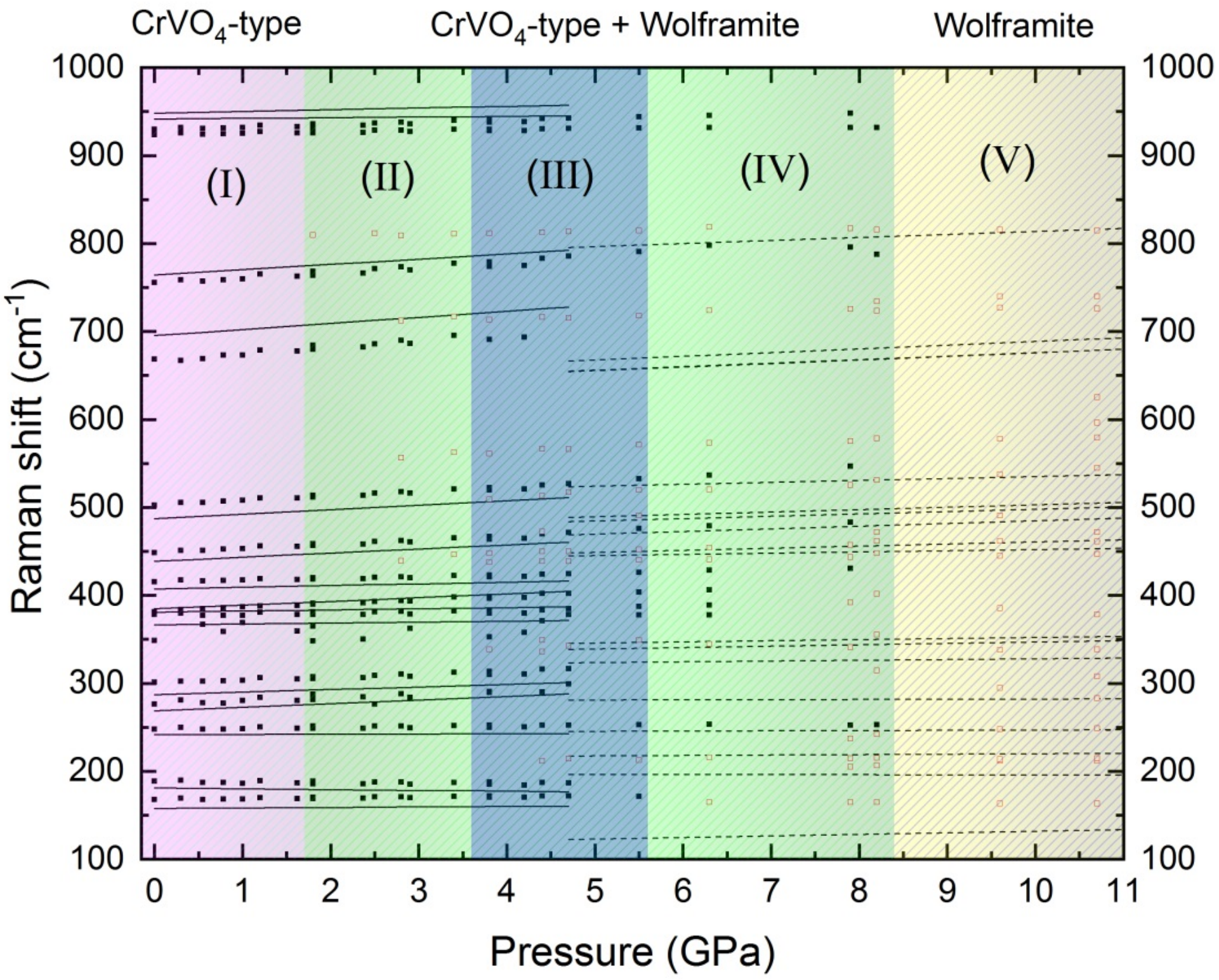

**Fig. 10.**

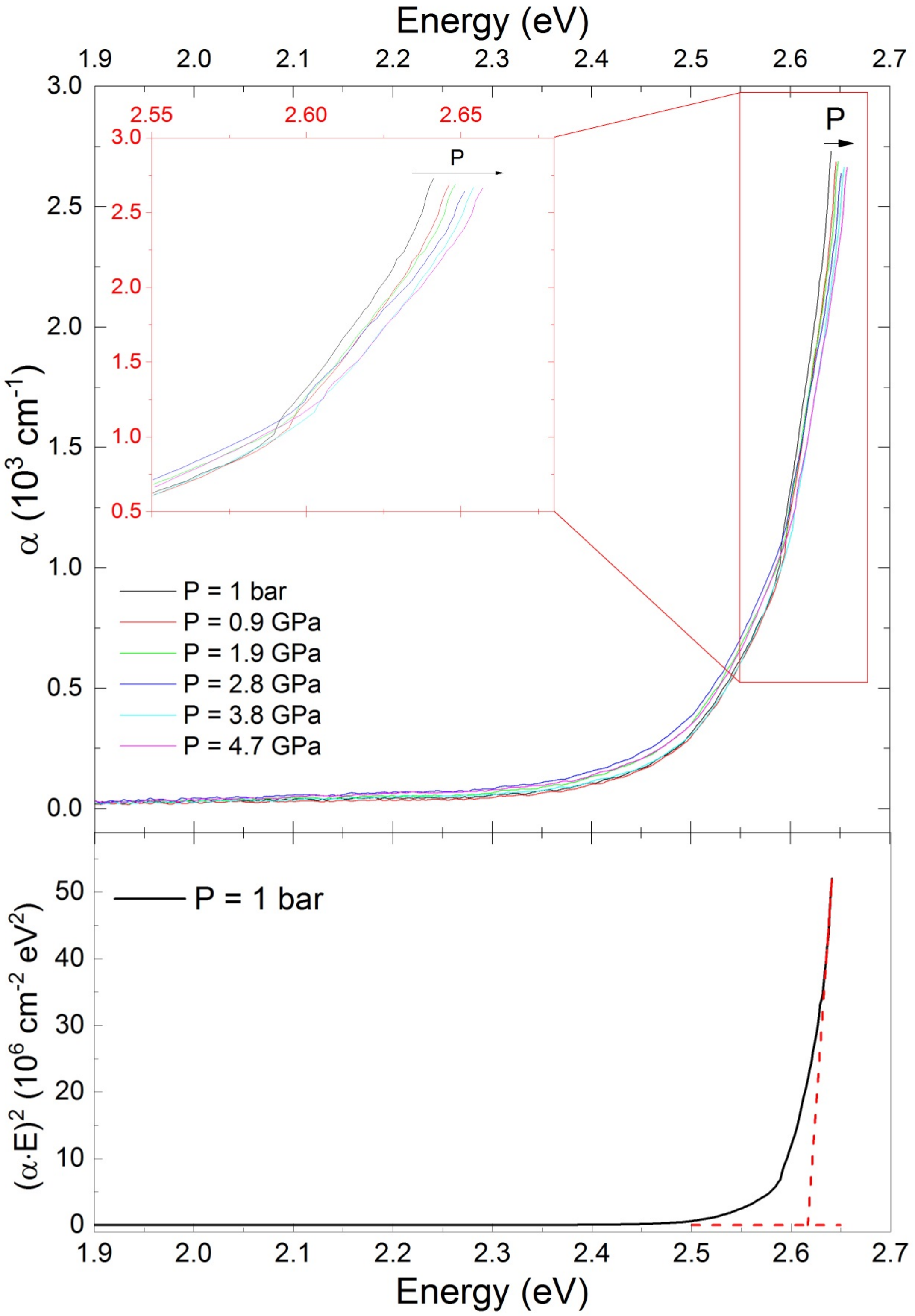

**Fig. 11.**

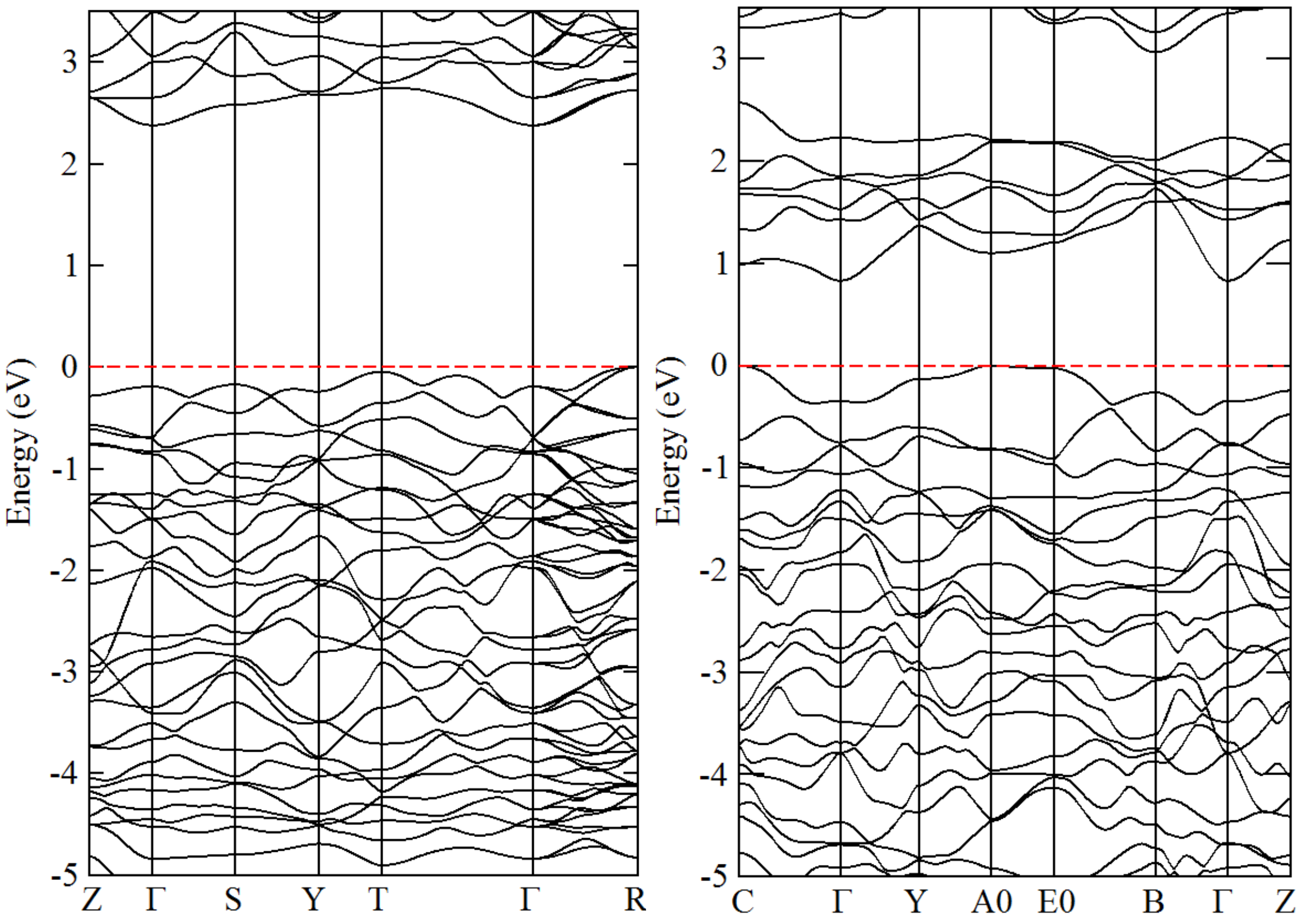

(a) (b)

**Fig. 12.**

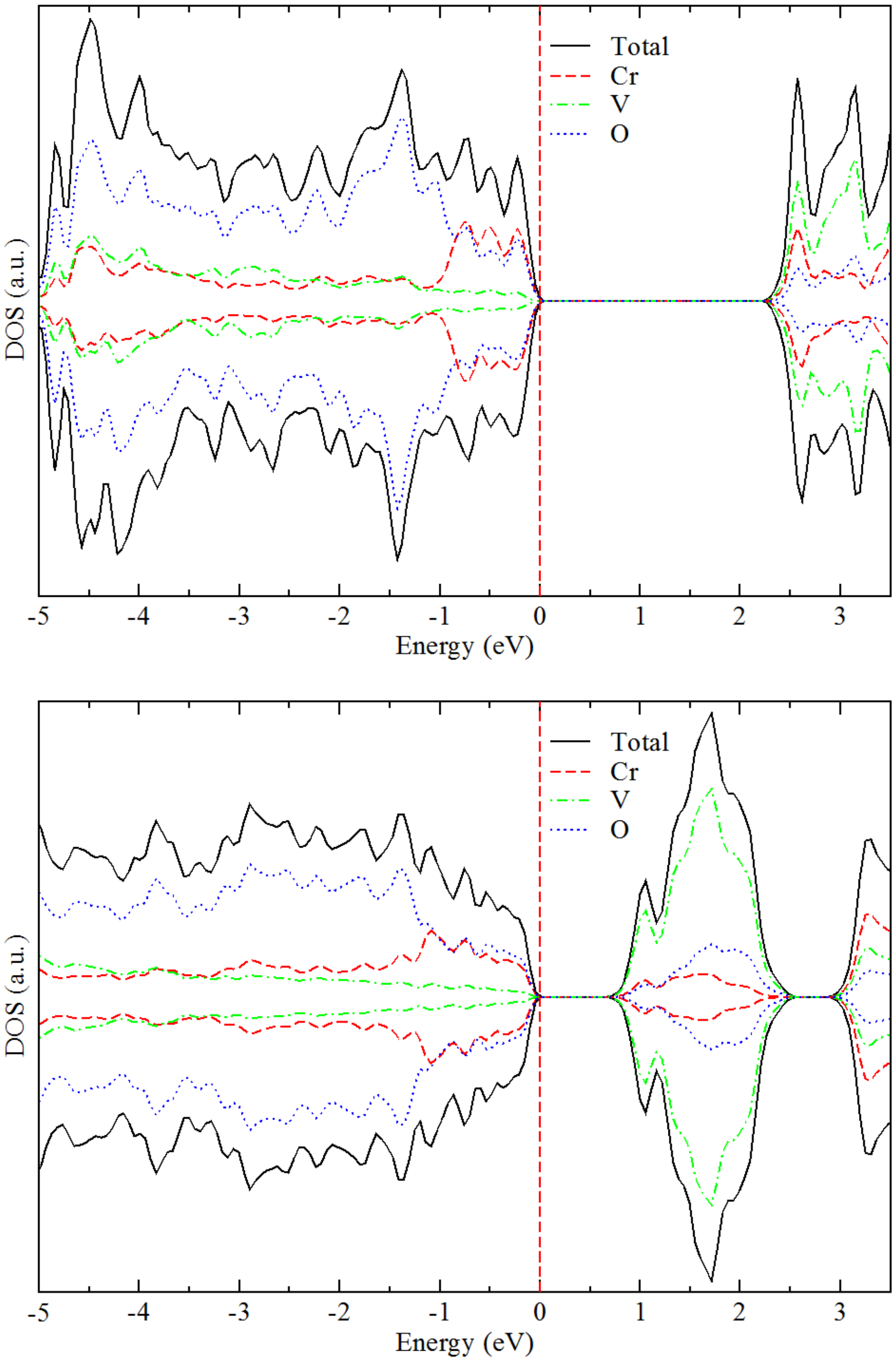

**Fig. 13.**

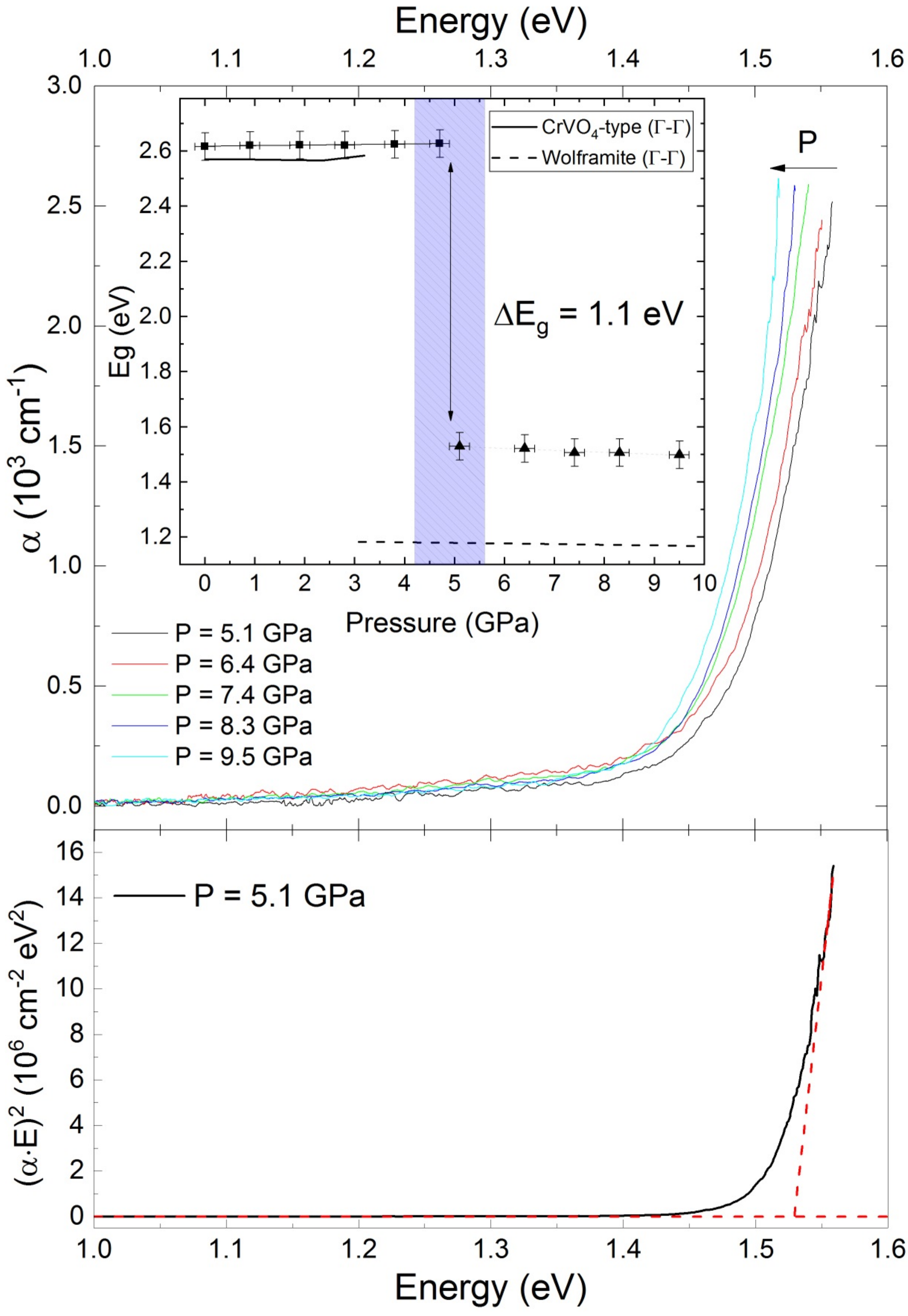

**Fig. 14.**

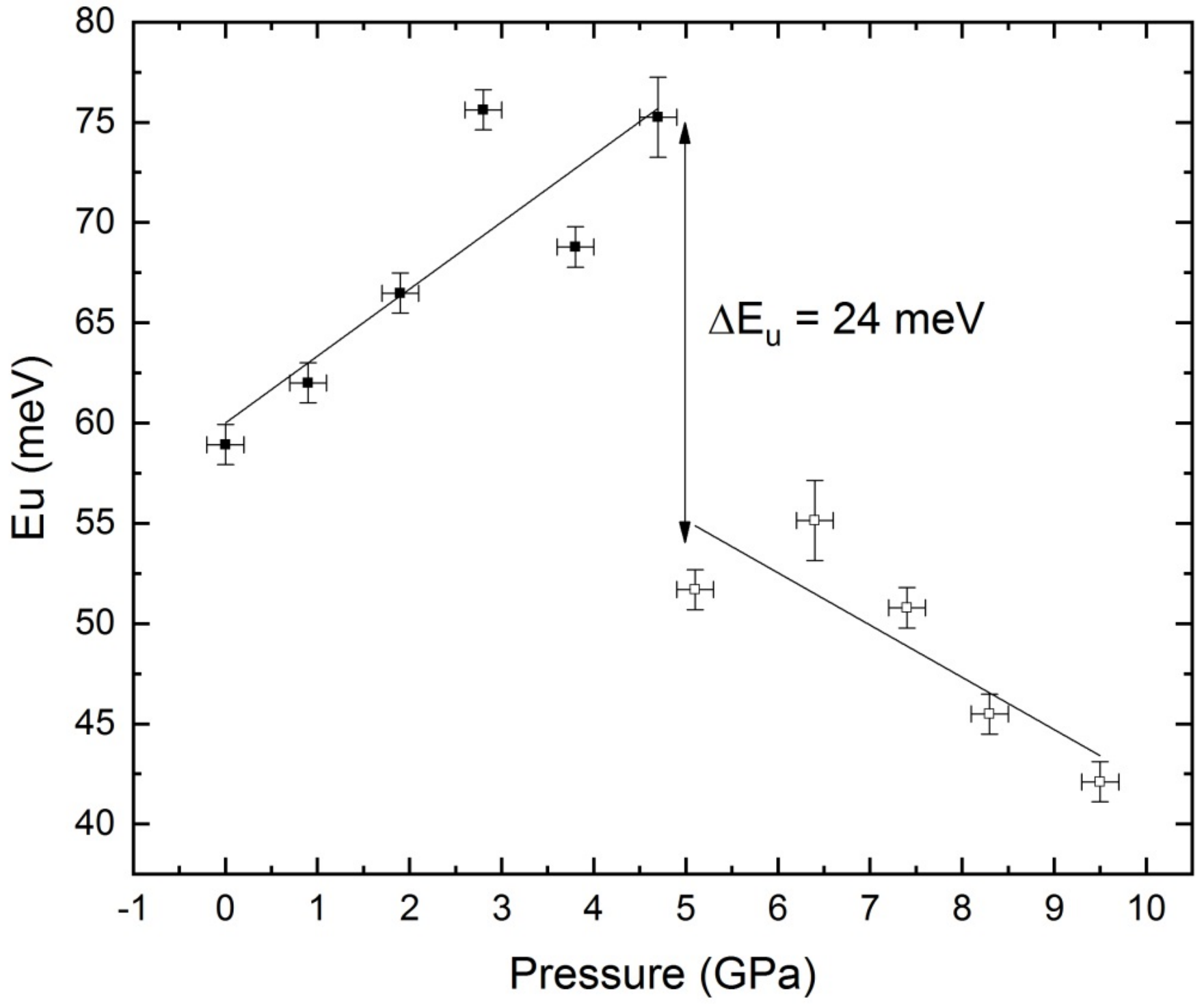

**Fig. 15.**

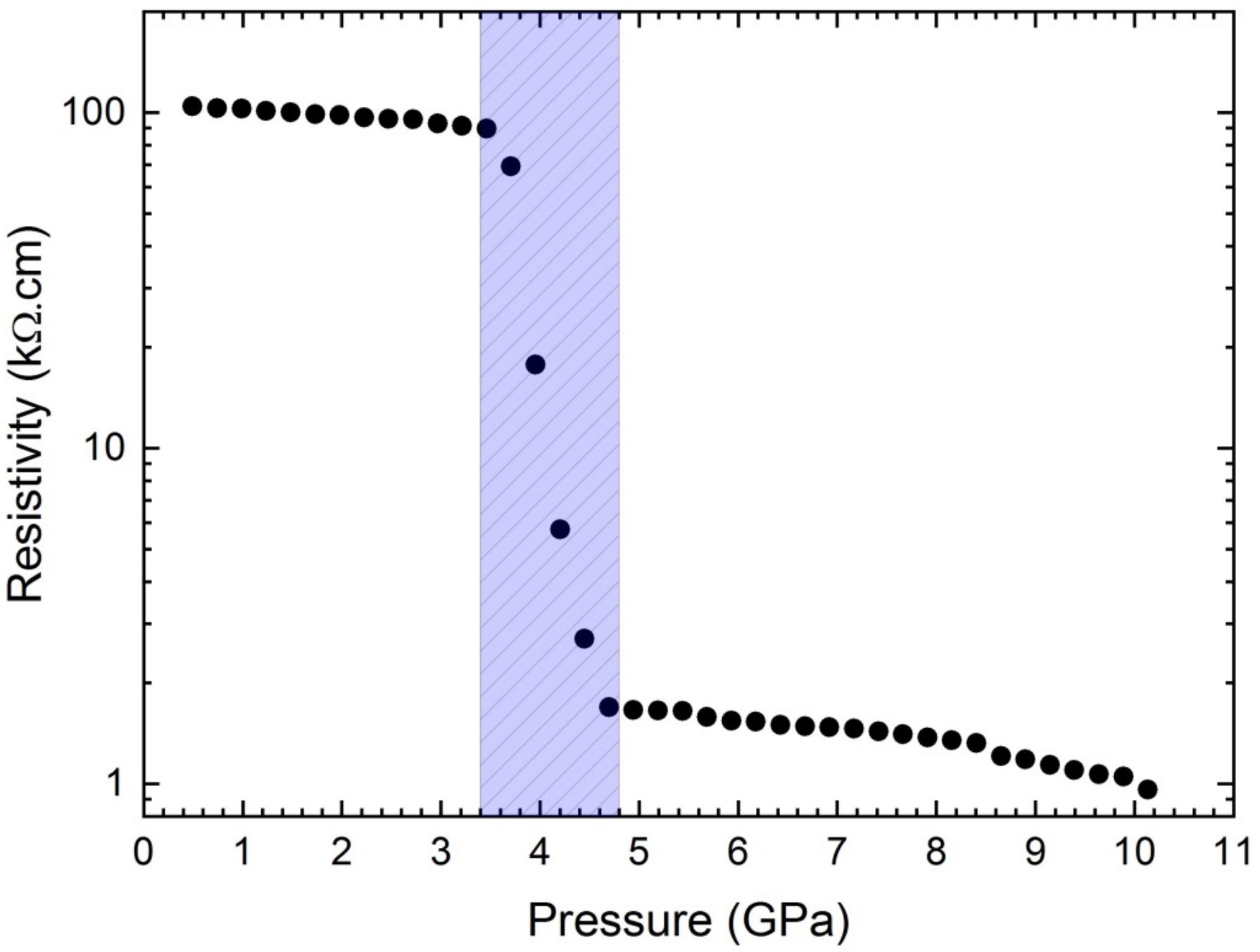

## TABLE CAPTIONS

**Table 1.** Unit-cell parameters and volume of the HP wolframite-type structure of $CrVO_4$, $InVO_4$ and $FeVO_4$ and bulk modulus and its pressure derivative at ambient pressure.

**Table 2.** Zero-pressure Raman phonon frequencies, pressure coefficients and Grüneisen parameters $\left(\gamma = \frac{B_0}{\omega_0}\frac{\partial\omega}{\partial P}\right)$ of $CrVO_4$-III. For the calculation of the Grüneisen parameters we have used the experimental bulk modulus. Assignment of $\nu_1$ and $\nu_3$ ($\nu_2$ and $\nu_4$): stretching (bending), T: translation and R: Rotation modes to the $VO_4$ tetrahedron from reference [66].

**Table 3.** Raman phonon frequencies and pressure dependence of wolframite-type phase at 7.26 GPa.

**Table 4.** Experimental and theoretical values of the energy band gap (Eg) and pressure coefficients for the LP and HP phase of $CrVO_4$.

**Table 5.** Experimental values of the Urbach energy (Eu) and pressure coefficients for the LP and HP phase of $CrVO_4$.

## TABLES

### Table 1.

| Method | CrVO$_4$ DFT | CrVO$_4$ Exp. | InVO$_4$ DFT [28] | InVO$_4$ DFT [27] | InVO$_4$ Exp. [26] | FeVO$_4$ DFT [20] | FeVO$_4$ Exp. [20] | FeVO$_4$ Exp. [62] |
|---|---|---|---|---|---|---|---|---|
| | [This work] | | | | | | | |
| Pressure (GPa) | 7.3 | 7 | 8.5 | 6.0 | 8.2 | 12.4 | 12.3 | ambient[a] |
| a (Å) | 4.37720 | 4.367(3) | 4.7009 | 4.776 | 4.714(5) | 4.4561 | 4.3952(8) | 4.511(2) |
| b (Å) | 5.42172 | 5.403(5) | 5.5197 | 5.588 | 5.459(6) | 5.4458 | 5.4361(8) | 5.527(2) |
| c (Å) | 4.76152 | 4.703(3) | 4.8849 | 4.958 | 4.903(5) | 4.7761 | 4.7819(7) | 4.851(2) |
| β | 90.153 | 90.27(8) | 92.62 | 92.89 | 93.8(3) | 90.493 | 89.87(2) | 90.867(2) |
| V (Å$^3$) | 113.0 | 115.5(6) | 126.62 | 132.17 | 125.89(2) | 115.9 | 114.25(3) | |
| $B_0$ (GPa) | 219.4 | 146(12) | 166.1 | 183.0 | 168(9) | 176.8 | 174(8) | |
| $B_0$' | 4.0 | 4.0 | 4.26 | 6.0 | 4.0 | 5.4 | 4.0 | |

**a)** Recovered sample after HTHP synthesis.

### Table 2.

| Mode | ω (cm$^{-1}$) Theo. | ω (cm$^{-1}$) Exp. (±2) | dω/dP (cm$^{-1}$/GPa) Theo. | dω/dP (cm$^{-1}$/GPa) Exp. | γ Theo. | γ Exp. | Assignment Theo. |
|---|---|---|---|---|---|---|---|
| | **CrVO$_4$-type** | | | | | | |
| $\omega_1$ | 158 | 168 | 0.59 | 0.8(1) | 0.37 | 0.47(8) | T($B_{3g}$) |
| $\omega_2$ | 181 | 189 | -0.88 | -0.5(2) | -0.49 | -0.2(1) | T($B_{1g}$) |
| $\omega_3$ | 242 | 248 | 0.24 | 0.8(1) | 0.10 | 0.33(5) | T($A_g$) |
| $\omega_4$ | 269 | 277 | 4.09 | 2.9(6) | 1.47 | 1.0(2) | R($B_{1g}$) |
| $\omega_5$ | 287 | 301 | 2.91 | 3.0(2) | 0.99 | 0.99(7) | R($B_{2g}$) |
| $\omega_6$ | 367 | 349 | 1.00 | 0.6(1.7) | 0.27 | 0.2(5) | $\nu_2$($A_g$) |
| $\omega_7$ | 381 | 378 | 1.30 | 1.7(2) | 0.34 | 0.44(5) | R($B_{1g}$) |
| $\omega_8$ | 385 | 381 | 4.27 | 4.0(1) | 1.08 | 1.03(4) | $\nu_4$($B_{3g}$) |
| $\omega_9$ | 408 | 416 | 1.85 | 1.9(1) | 0.45 | 0.45(3) | $\nu_2$($A_g$) |
| $\omega_{10}$ | 439 | 449 | 4.61 | 4.5(1) | 1.02 | 0.99(4) | $\nu_4$($B_{2g}$) |
| $\omega_{11}$ | 487 | 502 | 5.06 | 5.3(1) | 1.01 | 1.04(3) | $\nu_4$($B_{3g}$) |
| $\omega_{12}$ | 695 | 669 | 6.93 | 6.8(5) | 0.97 | 1.01(8) | $\nu_3$($B_{1g}$) |
| $\omega_{13}$ | 765 | 756 | 5.90 | 6.4(3) | 0.75 | 0.84(4) | $\nu_3$($A_g$) |
| $\omega_{14}$ | 942 | 924 | 0.78 | 1.0(1) | 0.08 | 0.11(1) | $\nu_1$($A_g$) |
| $\omega_{15}$ | 948 | 930 | 1.93 | 2.4(1) | 0.20 | 0.26(1) | $\nu_3$($B_{3g}$) |

**Table 3.**

| Mode | ω (cm$^{-1}$) Theo. | Exp. (±5) | dω/dP (cm$^{-1}$/GPa) Theo. | Exp. | Assignment Theo. |
|---|---|---|---|---|---|
| $\omega_1$ | 133 | 163 | 1.77 | -0.5(1) | $B_g$ |
| $\omega_2$ | 196 | 212 | -0.14 | 2.4(7) | $A_g$ |
| $\omega_3$ | 220 | 215 | 0.48 | 0.3(1) | $B_g$ |
| $\omega_4$ | 247 | 249 | 0.27 | 4(1) | $B_g$ |
| $\omega_5$ | 282 | 283 | 0.21 | -- | $B_g$ |
| $\omega_6$ | 329 | 308 | 0.84 | -- | $A_g$ |
| $\omega_7$ | 348 | 338 | 1.60 | -- | $A_g$ |
| $\omega_8$ | 353 | 378 | 1.25 | -- | $B_g$ |
| $\omega_9$ | 453 | 447 | 1.39 | 1.4(2) | $A_g$ |
| $\omega_{10}$ | 463 | 461 | 2.40 | 2.6(3) | $B_g$ |
| $\omega_{11}$ | 487 | 472 | 3.03 | -- | $A_g$ |
| $\omega_{12}$ | 500 | 545 | 2.66 | 4.7(3) | $B_g$ |
| $\omega_{13}$ | 505 | 579 | 2.73 | 2.8(3) | $B_g$ |
| $\omega_{14}$ | 537 | 596 | 2.21 | -- | $A_g$ |
| $\omega_{15}$ | 679 | 625 | 4.04 | -- | $B_g$ |
| $\omega_{16}$ | 679 | 726 | 3.97 | 0.9(9) | $A_g$ |
| $\omega_{17}$ | 692 | 740 | 4.19 | 3.9(3) | $B_g$ |
| $\omega_{18}$ | 816 | 815 | 3.42 | 0.8(1) | $A_g$ |

**Table 4.**

| Study type | LP phase $E_g$ (eV) | LP phase $dE_g/dP$ (meV/GPa) | LP phase Gap nature | $\Delta E_g$ collapse (eV) | HP phase $E_g$ (eV) | HP phase $dE_g/dP$ (meV/GPa) | HP phase Gap nature | *k* points |
|---|---|---|---|---|---|---|---|---|
| Theoretical | 2.57 | 4.1 | Direct | 1.4 | 1.18 | -2.77 | Direct | $\Gamma \rightarrow \Gamma$ |
| Experimental | 2.62(5) | 1.9(3) | Direct | 1.1(1) | 1.53(5) | -7.1(1.1) | Direct | |

The band-gap collapse at the phase transition (ΔEg) is also given.

**Table 5.**

| HP phase $E_u$ (meV) | $dE_u/dP$ (meV/GPa) | $\Delta E_u$ collapse (meV) | $E_u$ (meV) | $dE_u/dP$ (meV/GPa) |
|---|---|---|---|---|
| 59(1) | 3.3(9) | 24(1) | 52(1) | -2.6(9) |